%% file: rome.tex
\newif\ifextended\extendedtrue
\newif\ifaec\aecfalse
\newif\ifreview\reviewfalse

\PassOptionsToPackage{svgnames,dvipsnames}{xcolor}
\PassOptionsToPackage{unicode}{hyperref}
\PassOptionsToPackage{naturalnames}{hyperref}

\ifreview
\documentclass[authoryear,acmsmall,screen,anonymous,review]{acmart}
\else
\documentclass[authoryear,acmsmall,screen]{acmart}
\fi

\usepackage{rome}

\newcommand\Appendix[1]{\ifreview the appendices, included in the anonymized supplemental material\else \cref{#1}, included in the extended version\fi\xspace}
\newcommand\Supplemental{\ifreview, included in the anonymized supplementary material\else \citep{artifactRome}\fi\xspace}

\title{Abstracting Extensible Recursive Functions}

\author{Alex Hubers}
\orcid{0000-0002-6237-3326}
\affiliation{
  \department{Department of Computer Science}
  \institution{The University of Iowa}
  \streetaddress{14 MacLean Hall}
  \city{Iowa City}
  \state{Iowa}
  \country{USA}}
\email{alexander-hubers@uiowa.edu}

\author{Apoorv Ingle}
\orcid{0000-0002-7399-9762}
\affiliation{
  \department{Department of Computer Science}
  \institution{The University of Iowa}
  \streetaddress{14 MacLean Hall}
  \city{Iowa City}
  \state{Iowa}
  \country{USA}}
\email{garrett-morris@uiowa.edu}

\author{Andrew Marmaduke}
\orcid{0000-0002-5230-6728}
\affiliation{
  \department{Department of Computer Science}
  \institution{The University of Iowa}
  \streetaddress{14 MacLean Hall}
  \city{Iowa City}
  \state{Iowa}
  \country{USA}}
\email{garrett-morris@uiowa.edu}

\author{J. Garrett Morris}
\orcid{0000-0002-3992-1080}
\affiliation{
  \department{Department of Computer Science}
  \institution{The University of Iowa}
  \streetaddress{14 MacLean Hall}
  \city{Iowa City}
  \state{Iowa}
  \country{USA}}
\email{garrett-morris@uiowa.edu}

\begin{document}

\begin{abstract}
  We explore recursive programming with extensible data types. Row types make the structure of data types first class, and can express a variety of type system features including record subtyping and combination of case branches. Our goal is the modular combination of recursive types and of recursive functions over them. The most significant challenge is in recursive function calls, which may need to account for new cases in a combined type. We introduce \defemph{extensible histomorphisms}, Mendler-style descriptions of recursive functions in which recursive calls can happen at larger types, and show that they provide expressive recursion over extensible data types. We formalize our approach in \Rome, a row type theory with support for recursive terms and types.
\end{abstract}

\maketitle

\section{Introduction}

The expression problem \citep{Wadler98} is (now) an old name for an older problem \citep{Reynolds75}: to extend the definition (by cases) of a recursive data type, both by adding new cases to the data type and by adding new operations over it, without compromising either type safety or separate compilation. This problem has motivated both dedicated language features for extensibility, such as row typing \citep{Wand87,Remy89,HarperP91,Chlipala10}, and encodings using features like type classes \citep{Swierstra08} and generic programming~\cite{Wadler98}.

This paper explores a natural extension of the expression problem. Our goals are, first, to decompose a recursive data type into modular (that is, independent) cases and define operations recursively over those cases, and second, to recover the original data type and the operations over it by combining the cases. To achieve this goal, we will define a new row type theory, \Rome, which extends the row type theory \RO, with recursive types and more expressive row type constructors, and then define \emph{extensible histomorphisms}, a new encoding of traversals, in \Rome.

\subsubsection*{An extended expression problem}

Programming languages can frequently be decomposed into independent sublanguages. Even PCF~\citep{Plotkin77} could be regarded as the combination of a language for functions (variables, abstractions, applications), a language for Booleans (constants, conditionals), a language for arithmetic, and a language of recursive definition. These sublanguages could be recombined in many different ways---indeed, one could see much research in programming languages from exactly this perspective. We could add a sublanguage for explicit polymorphism, and remove the languages for arithmetic and Booleans. Or, we could add a language for data types---with or without explicit polymorphism. We might replace the arbitrary fixed point operator with structured recursion schemes, relying on either arithmetic or data types. And so forth.

The challenge is to implement these languages in a way that is as modular as their informal description. There are several aspects to this challenge. We describe these informally now, but demonstrate them concretely later in the paper~\cref{sec:xr}.
\begin{itemize}[leftmargin=\parindent]
\item Individual languages (that is to say, data types) should be defined independently and support arbitrary (recursive) combination. We should be able to define the languages of Boolean, arithmetic, and functional expressions independently, and define terms in each of their intersections. For example, in defining the negation function, we should not have to commit to the presence or absence of arithmetic expressions, nor in defining the successor function to the presence or absence of Boolean expressions. In contrast, most existing row type theories define rows only by extension, which would rule out at least one of the three possible intersections.
\item Computations, like evaluation, should also be defined independently, and support arbitrary combination. This problem has two aspects. First, we cannot just combine functions at the top level. In a combined language, functional expressions (for instance, an application) may have subexpression from other languages (for instance, a number). Similarly, recursive calls in the combined evaluator must be to the combined evaluator, not to its individual parts. Second, results types may vary. Evaluating arithmetic expressions might produce natural numbers, while evaluating Boolean expressions produces Boolean values, and the results of evaluating a function---that is to say, the body of a closure---may vary with the input language.
\item Transformations, such as substitution or desugaring, should be defined modularly and support generic extension. The arithmetic and Boolean languages, for example, do not contain variables, and so do not interact with substitution in an interesting way. It should be possible to extend substitution to these languages without manually enumerating and traversing each case.
\end{itemize}

\noindent
These challenges are difficult to meet with current approaches to extensible data types. The vast majority of row type systems \citep{Wand87,Remy89,PottierR05,BlumeAC06,Leijen05,Garrigue98} express rows by extension, rather than concatenation. Thus, while they could express the extension of, say, \lstinline!FunE! to include arithmetic expressions, they would not support the independent development of \lstinline!FunE! and \lstinline!ArithE!. There are several row type systems with concatenation \citep{HarperP91,Remy92,Chlipala10}, but they do not address combination of recursive functions. OCaml includes a flexible system of polymorphic variants with subtyping \citep{Garrigue00,Garrigue10}. While polymorphic variants are quite expressive, they can lose precision in typing, and still require some enumeration of cases to describe generic traversals. Finally, there are encodings of extensible recursive types with concatenation \citep{Swierstra08}, including some that support combining functions over arbitrary rows \citep{Bahr14,Morris15}. However, these systems rely on expressing recursion via folds, which limits their expressiveness.

\subsubsection*{The \Rome type theory}

We propose an approach to expressing extensible recursive types and recursive functions that meets all four of these challenges. We work in \Rome, an extension of the row type system \RO \citep{MorrisM19,HubersM23}. \RO differs from most other row type systems in three ways: rows are constructed by concatenating arbitrary rows, rather than by concrete extension; the structure of row types is captured using containment and concatenation relations rather than directly in the grammar of types; and, label-generic programming primitives support transformations of extensible data types regardless of their particular constructors or field labels. We extend \RO in several ways. We add iso-recursive types and fixed point computations, in an entirely standard fashion. We introduce the \emph{relative complement} of one row with respect to another, which is essential for typing our encodings and significantly simplifies typing several existing operators in \RO. Finally, we have simplified and mechanized the metatheory of \Rome, including showing type normalization (by evaluation), decidability of type equality, and type soundness.

\subsubsection*{Extensible histomorphisms}

An expressive type theory, in and of itself, does not constitute a solution to the extended expression problem. We must also be able to encode recursive traversals in a way that both supports extensibility and is expressive enough for practical programming. Our solution is to abstract recursive traversals as \emph{extensible histomorphisms}.
Extensible histomorphisms are Mendler-style \citep{Mendler91} encodings of histomorphic recursive functions. By keeping recursive subdata abstract, we are able to guarantee that recursive calls will still make sense when combining extensible histomorphisms. Keeping recursive subdata completely abstract, however, would rule out many useful functions. Extensible histomorphisms use \Rome containment predicates to provided a lower bound for the structure of subdata. Finally, unlike other recursion schemes, we define the output of a extensible histomorphism as a function of its input. We will show that extensible histomorphisms can express interesting, extensible evaluation functions, addressing the sublanguage expression problem.

\subsubsection*{Contributions}

In summary, this paper contributes:
\begin{itemize}
\item The \Rome type theory~\cref{sec:rome}, which extends existing concatenative row type theories with a novel row complement operator, recursion, and recursive types;
\item extensible histomorphisms~\cref{sec:xr}, an abstraction for extensible recursive functions, along with examples of their expressiveness; and
\item the metatheory of \Rome~\cref{sec:metatheory}, given by a type normalization algorithm, operational semantics, and the expected proofs of type preservation and progress.
\end{itemize}
We conclude by discussing related~\cref{sec:related} and future~\cref{sec:future} work. We have mechanized the metatheory of \Rome and its essential properties using the Agda interactive theorem prover, and have built an experimental implementation of \Rome that typechecks and evaluates all the examples in this paper.

\section{The \Rome Calculus}
\label{sec:rome}

This section defines the \Rome calculus. We begin by formally describing its syntax~\cref{sec:syntax} and type system~\cref{sec:typing}. Operations on rows in \Rome are captured by predicates in types, which are given meaning by its entailment relation~\cref{sec:entailment}. We then give a series of examples illustrating the features and expressiveness of \Rome: its approach to extensible records and variants~\cref{sec:variants-and-records}, its support for first-class labels~\cref{sec:first-class-labels}, its label generic operators~\cref{sec:label-generic-programming}, and finally the utility of relative row complements~\cref{sec:row-complements}.

\subsection{Overview}
\label{sec:rome-overview}
\InlineOn

In a row type system~\citep{Wand87,Remy89}, rows are first-class representations of the structure of data, typically written as lists of label-type associations. Data types, such as records and variants, are built from rows. Extensibility arises from row polymorphism. For example, the function that selects label !x! with type $Int$ from a record might have a type $\forall r. \Pi \Row { \LabTy {\Lab x} {\Int} \mid r} \to \Int$. Here, $r$ can be instantiated with any row, so this function applies equally well to two-dimensional points $\Pi \Row { \LabTy {\Lab x} {\Int}, \LabTy {\Lab y} {\Int}}$ and function abstractions $\Pi \Row { {\LabTy {\Lab x} {Int}}, {\LabTy {\Lab {body}} {Expr}}}$. Rows are typically required to have all distinct labels, enforced by kinding, and to be equivalent up to reordering.

We build on the row type system \Rose, initially proposed by \citet{MorrisM19}, and extended as \RO by \citet{HubersM23}. \Rose differs from most other row type systems in two ways. Most row type systems describe rows by extension: the row $\Row { \LabTy {\Lab x} {\Int} \mid r}$ extends row $r$ with the association $\LabTy {\Lab x} \Int$. \Rose, like the type systems of \citet{HarperP91} and \citet{Chlipala15a}, permits arbitrary concatenation of rows. Unlike those systems, \Rose describes rows using \emph{qualified types}. Rather than the concatenation of rows $\rho_1$ and $\rho_2$ being represented directly as a type $\rho_1 \odot \rho_2$, it is treated as a three-place predicate on types $\rho_1 \odot \rho_2 \sim \rho_3$. Qualified type systems are proof relevant: the proof that $\PlusP {\rho_1} {\rho_2} {\rho_3}$ will determine the behavior of the primitive record and variant operators. \Rose's approach has several advantages. One is expressiveness: examples that are difficult to express in most row type systems can be simple to express in \Rose. Another is flexibility: By adjusting the meaning of the predicates, \Rose can capture a variety of different models of rows, such as those that allow duplicate labels. In this paper, we will limit ourselves to one model of rows, in which labels are required to be unique and entries in rows may be freely permuted.

\InlineOff
\subsection{Syntax}
\label{sec:syntax}

\begin{figure}
\begin{gather*}
\begin{array}{r@{\hspace{7px}}l@{\qquad\qquad}r@{\hspace{7px}}l}
  \text{Term variables} & x \in \mathcal X &
  \text{Type environments} & \Gamma ::= \varepsilon \mid \Gamma, x : \tau \\
  \text{Type variables} & \alpha \in \mathcal A &
  \text{Predicate environments} & \Phi ::= \varepsilon \mid \Phi, v : \pi \\
  \text{Labels} & \ell \in \mathcal L &
  \text{Kind environments} & \Delta ::= \varepsilon \mid \Delta, \alpha : \kappa
\end{array}
\\[5px]
\begin{doublesyntaxarray}
  \mcl{\text{Kinds}} & \kappa & ::= & \TypeK \mid \LabK \mid \RowK \kappa \mid \kappa \to \kappa \\
  \mcl{\text{Predicates}} & \pi, \psi & ::= & \LeqP \rho \rho \mid \PlusP \rho \rho \rho \\
  \text{Types} & \mcr{\Types \ni \phi, \tau, \upsilon, \rho, \xi} & ::= & \alpha \mid T \mid \pi \then \tau \mid \forall \alpha\co\kappa. \tau \mid \lambda \alpha\co\kappa. \tau \mid \tau \, \tau \\
  & & & \mid & \RowIx i 0 m {\LabTy {\xi_i} {\tau_i}} \mid \ell \mid \Sing \tau \mid \Map \phi \mid \rho \Compl \rho \\
  \mcl{\text{Type constants}} & T & ::= & (\to) \mid \KFam \Pi \kappa \mid \KFam \Sigma \kappa \mid \mu \\
  \text{Terms} & \mcr{\Terms \ni M, N} & ::= & x \mid K \mid \lambda x \co \tau. M \mid M \, N \mid \Lambda \alpha \co \kappa. M \mid \AppT M \tau \mid
    \Lambda v \co \pi. M \mid \AppT M Q \\
  & & & \mid & \Sing\tau \mid \LabTermX \Xi M N \mid \UnlabelX \Xi M N \qquad \qquad \Xi \in \Set {\Pi, \Sigma} \\
  \mcl{\text{Term constants}} & K & ::= & \Prj {} \mid (\Concat) \mid \Inj \mid (\Branch) \mid \KFam \Syn \kappa \mid \KFam \Ana \kappa \mid \In \mid \Out \mid \Fix
\end{doublesyntaxarray}
\end{gather*}
\caption{Syntax}
\label{fig:syntax-types-terms}
\end{figure}

Syntax for the types and terms of \Rome is given in \cref{fig:syntax-types-terms}. Syntax for evidence variables $v$ and terms $Q$ will be given shortly~\cref{sec:entailment}.

Labels (i.e., record field and variant constructor names) live at the type level, and are classified by kind $\LabK$. Rows of kind $\kappa$ are classified by $\RowK \kappa$.

When possible, we use $\phi$ for type functions, $\rho$ for row types, and $\xi$ for label types. Singleton types $\Sing \tau$ are used to make (otherwise implicit) type parameterization explicit.  $\Map\phi$ maps type operator $\phi$ across a row. In practice, we leave the map operator implicit, using kind information to infer the presence of maps. We define a families of  $\Pi$ and $\Sigma$ constructors, describing record and variants at various kinds; in practice, we can determine the kind annotation from context. Finally, $\mu$ builds isorecursive types.

The type $\pi \then \tau$ denotes a qualified type. In essence, the predicate $\pi$ restricts the instantiation of the type variables in $\tau$. Our predicates capture relationships among rows: $\LeqP {\rho_1} {\rho_2}$ means that $\rho_1$ is \emph{contained} in $\rho_2$, and $\PlusP {\rho_1} {\rho_2} {\rho_3}$ means that $\rho_1$ and $\rho_2$ can be \emph{combined} to give $\rho_3$.

Row literals are sequences of labeled types $\LabTy {\xi_i} {\tau_i}$. We write $0 \dots m$ to denote the set of naturals up to (but not including) $m$. We will frequently use $\varepsilon$ to denote the empty row. Occasionally we will need to treat rows as lists; in that case, we will write $\Row {\LabTy \xi \tau, \rho}$ for the row that begins with the labeled type $\LabTy \xi \tau$ and continues with $\rho$. This is purely metatheoretic notation; we do not have row extension as a part of the concrete syntax of our language.

Much of the interest in terms is deferred to the constants $K$. Type abstraction $\Lambda \alpha\co\kappa. M$ and application $\AppT M \tau$ are expected to be implicit in source programs, and evidence abstraction $\Lambda v\co\pi. M$ and application $M \AppE Q$ are always implicit. We pun $\Sing \tau$ for the unique value of type $\Sing \tau$. The terms $\LabTermX \Xi M N$ and $\UnlabelX \Xi M N$ construct and deconstruct singleton records (with $\Xi = \Pi$) and variants ($\Xi = \Sigma$). These forms are overloaded in \RO, and indeed in our implementation, but we make the distinction explicit in the formal account for clarity.  We do not include kind polymorphism, as it is an orthogonal concern, and so treat $\KFam \Syn \kappa$ and $\KFam \Ana \kappa$ as kind-indexed families of terms.

We track separate environments for type ($\Delta$), evidence ($\Phi$), and term ($\Gamma$) variables.

\subsection{Typing}
\label{sec:typing}

\begin{figure}
\begin{small}
\begin{gather*}
\fbox{$\KindJ \Delta \tau \kappa$} \;\; \fbox{$\PredJ \Delta \pi$}
\\
\ib{\irule[\krule{var}]
          {\alpha : \kappa \in \Delta};
          {\Delta \vdash \alpha : \kappa}}
\rsp
\ib{\irule[\krule{T}]
          {T : \kappa};
          {\KindJ \Delta T \kappa}}
\rsp
\ib{\irule[\krrule{lab}]
          { };
          {\KindJ \Delta \ell \LabK}}
\rsp
\ib{\irule[\krule{$\then$}]
          {\PredJ \Delta \pi}
          {\KindJ \Delta \tau \TypeK};
          {\KindJ \Delta {\pi \then \tau} \TypeK}}
\\
\ib{\irule[\krule{$\forall$}]
          {\KindJ {\Delta, \alpha : \kappa} \tau \TypeK};
          {\KindJ \Delta {\forall \alpha\co\kappa. \tau} \TypeK}}
\rsp
\ib{\irule[\krule{$\I\to$}]
          {\KindJ {\Delta, \alpha : \kappa_1} \tau \kappa_2};
          {\KindJ \Delta {\lambda \alpha \co \kappa_1. \tau} {\kappa_1 \to \kappa_2}}}
\rsp
\ib{\irule[\krule{$\E\to$}]
          {\KindJ \Delta {\tau_1} {\kappa_1 \to \kappa_2}}
          {\KindJ \Delta {\tau_2} {\kappa_1}};
          {\KindJ \Delta {\tau_1 \, \tau_2} {\kappa_2}}}
\\
\ib{\irule[\krrule{sing}]
          {\KindJ \Delta \xi \kappa};
          {\KindJ \Delta {\Sing\xi} \TypeK}}
\rsp
\ib{\irule[\krrule{map}]
          {\KindJ \Delta \phi {\kappa_1 \to \kappa_2}};
          {\KindJ \Delta {\Map\phi} {\RowK{\kappa_1} \to \RowK{\kappa_2}}}}
\rsp
\ib{\irule[\ksrule{row}]
          {\KindJ \Delta {\xi_i} \LabK}
          {\KindJ \Delta {\tau_i} \kappa}
          {\forall i < j. \, \xi_i \LLt \xi_j};
          {\KindJ \Delta {\RowIx i 0 n {\LabTy {\xi_i} {\tau_i}}} {\RowK \kappa}}}
\\
\ib{\irule[\krrule{compl}]
          {\KindJ \Delta {\rho_2} {\RowK\kappa}}
          {\KindJ \Delta {\rho_1} {\RowK\kappa}};
          {\KindJ \Delta {\rho_2 \Compl \rho_1} {\RowK\kappa}}}
\rsp
\ib{\irule[\krrule{$\lesssim$}]
          {\KindJ \Delta {\rho_i} {\RowK \kappa}};
          {\PredJ \Delta {\LeqP  {\rho_1} {\rho_2}}}}
\rsp
\ib{\irule[\krrule{$\odot$}]
          {\KindJ \Delta {\rho_i} {\RowK \kappa}};
          {\PredJ \Delta {\PlusP {\rho_1} {\rho_2} {\rho_3}}}}
\\[5pt]
\fbox{$T : \kappa$}
\\
\begin{aligned}
(\to) &: \TypeK \to \TypeK \to \TypeK \\
\mu &: (\TypeK \to \TypeK) \to \TypeK
\end{aligned}
\qquad\qquad
\begin{aligned}
\KFam \Pi \kappa &: \RowK \kappa \to \kappa \\
\KFam \Sigma \kappa &: \RowK \kappa \to \kappa \\
\text{where } &\kappa \neq \LabK
\end{aligned}
\end{gather*}
\end{small}
\caption{Kinding (types and predicates)}
\label{fig:kinding}
\end{figure}

\subsubsection*{Kinding}

Kinding is shown in \cref{fig:kinding}. The majority of rules should be unsurprising. The environment formation judgments are given in \Appendix{sec:extra-typing-rules}, and are also unsurprising.

In the kinding of row literals, we assume some strict ordering of label literals $\mathord < \subseteq \mathcal L \times \mathcal L$, and lift this to a relation on types $\LLt$ by defining
\begin{align*}
  \ell \LLt \ell' &\iff \ell < \ell' \\
  \tau\, \nLLt\, \tau' &\;\;\text{otherwise}
\end{align*}
We require that row literals be listed in ascending order. This has two consequences. First, because all types other than label literals are incomparable, a row literal containing a label variable must have exactly one entry. Second, we avoid spurious distinctions that might be introduced by having different literals represent the same row. Of course, the requirement that entries be listed in order can trivially be assured by type checking, and so need not be imposed upon a surface language.

The relative complement $\rho_{2} \Compl \rho_{1}$ is defined for arbitrary $\rho_1$ and $\rho_2$; however, it will only be useful when $\rho_1$ is contained in $\rho_2$. Constraining $\rho_2 \Compl \rho_1$ to only be well-kinded when $\LeqP {\rho_1} {\rho_2}$ would introduce unpleasant circularity in the kinding and entailment rules (described further in \secref{row-complements}).

We define families of $\Pi$ and $\Sigma$ constructors for arbitrary kinds $\kappa$. At kind $\star$, they describe record and variant types. At higher kinds, they describe constructors of record and variant types. This is not strictly necessary. Given a row of type constructors $\rho : \RowK{\TypeK\to\TypeK}$, for example, we could write $\lambda t. \Sigma (\Map {(\lambda f. f \, t)} \, \rho)$ instead of $\Sigma \rho$ (indeed, the former is the latter's normal form, which we describe in \secref{type-reduction}). We have found the cognitive load of overloading $\Pi$ and $\Sigma$ to be much less than the alternative syntactic overload. We disallow the formation of records and variants at label kind $\LabK$, which introduces metatheoretic challenges without any redeeming practical applications.

\subsubsection*{Type equivalence}
\newcommand\Subtract{\mathsf{subtract}}

\begin{figure}
\begin{small}
\begin{gather*}
\fbox{$\TEqvJ \Delta \tau \tau \kappa$} \; \; \fbox{$\PEqvJ \Delta \pi \pi$}
\\
\ib{\irule[\erule{$\beta$}]
          {\KindJ \Delta {(\lambda \alpha\co\kappa. \tau) \, \upsilon} {\kappa'}};
          {\TEqvJ \Delta {(\lambda \alpha\co\kappa. \tau)\,\upsilon} {\tau[\upsilon/\alpha]} {\kappa'}}}
\rsp
\begin{gathered}
\ib{\irule[\errule{lift$_\Xi$}]
          {\KindJ \Delta \rho {\RowK {\kappa \to \kappa'}}}
          {\KindJ \Delta \tau \kappa};
          {\TEqvJ \Delta {(\KFam \Xi {\kappa \to \kappa'} \rho) \, \tau} {\KFam \Xi {\kappa'} (\rho^\$ \, \tau)} {\kappa'}}!
          {\Xi \in \Set {\Pi, \Sigma}}}
\\
\text {where $\rho^\$ \, \tau = \Map {(\lambda f. f \, \tau)} \rho$}
\end{gathered}
\\
\ib{\irule[\erule{$\setminus$}]
          {\KindJ \Delta {\rho_i} {\RowK \kappa}};
          {\TEqvJ \Delta {\rho_2 \Compl \rho_1} {\Subtract \, \rho_2 \, \rho_1} {\RowK \kappa}}}
\rsp
\ib{\irule[\errule{map}]
          {\KindJ \Delta \phi {\kappa_1 \to \kappa_2}}
          {\KindJ \Delta {\RowIx i 0 n {\LabTy {\xi_i} {\tau_i}}} {\RowK{\kappa_1}}};
          {\TEqvJ \Delta {\Map\phi \, \RowIx i 0 n {\LabTy {\xi_i} {\tau_i}}} {\RowIx i 0 n {\LabTy {\xi_i} {\phi\,\tau_i}}} {\RowK{\kappa_2}}}}
\\
\ib{\irule[\errule{map$_\mathsf{id}$}]
          {\KindJ \Delta \rho {\RowK\kappa}};
          {\TEqvJ \Delta {\Map{(\lambda \alpha. \alpha)} \, \rho} \rho {\RowK\kappa}}}
\rsp
\begin{gathered}
\ib{\irule[\errule{map$_\circ$}]
          {\KindJ \Delta {\phi_1} {\kappa_2 \to \kappa_3}}
          {\KindJ \Delta {\phi_2} {\kappa_1 \to \kappa_2}}
          {\KindJ \Delta \rho {\RowK {\kappa_1}}};
          {\TEqvJ \Delta {\Map {\phi_1} \, (\Map {\phi_2} \, \rho)} {\Map {(\phi_1 \circ \phi_2)} \, \rho} {\kappa_3}}}
\\
\text{where $\phi_1 \circ \phi_2 = \lambda \alpha. \phi_1 \, (\phi_2 \, \alpha)$}
\end{gathered}
\\
\ib{\irule[\errule{map$_{\setminus}$}]
          {\KindJ \Delta {\phi} {\kappa_1 \to \kappa_2}}
          {\KindJ \Delta {\rho_i} {\RowK {\kappa_1}}};
          {\TEqvJ \Delta {\Map {\phi} \, (\rho_2 \setminus \rho_1)} {\Map \phi \, {\rho_2} \setminus \Map \phi \, {\rho_1}} {\kappa_2} }}
\rsp
\ib{\irule[\errule{$\Xi$}]
          {\KindJ \Delta \rho {\RowK {\RowK {\kappa}}}};
          {\TEqvJ \Delta {\KFam \Xi {\RowK \kappa} \, \rho} {\Map{{\KFam \Xi \kappa}}\, \rho} {\RowK {\kappa}}}!
          {\Xi \in \{ \Pi , \Sigma \}}}
\\[5pt]
\fbox{$\Subtract \, \rho \, \rho$}
\\
\begin{aligned}
  \Subtract \, \EmptyRow \, \rho &= \EmptyRow \\
  \Subtract \, \rho \, \EmptyRow &= \rho \\
  \Subtract \, \Row {\LabTy \ell \tau, \rho} \, \Row{\LabTy {\ell'} {\tau'}, {\rho'}} &=
    \begin{cases}
      \Subtract \, \rho \, \rho' &\text{if $\ell = \ell'$ and $\tau = \tau'$} \\
      \Row {\LabTy \ell \tau, \Subtract \, \rho \, \Row {\LabTy {\ell'} {\tau'}, \rho'}} &\text{if $\ell < \ell'$} \\
      \Subtract \, \Row {\LabTy \ell \tau, \rho} \, \rho' &\text{if $\ell > \ell'$}
    \end{cases}
\end{aligned}
\end{gather*}
\end{small}
\caption{Type equivalence}
\label{fig:equivalence}
\end{figure}

The interesting cases of type equivalence are shown in \cref{fig:equivalence}.  The relative complement operator for row literals is implemented by the function $\Subtract \, \rho_{2} \, \rho_1$, which in turn relies on the ordering of rows. Rule \errule{map$_{\setminus}$} commutes mapping over row complements; this rule is derivable when $\rho_2$ and $\rho_1$ are literals. Rule \errule{map} implements the map operator for row literals. Rules \errule{map$_\mathsf{id}$} and \errule{map$_\circ$} observe that $\Map -$ behaves as a functor map for rows. Of course, these rules follow from \errule{map} in the case where $\rho$ is a row literal. Rules \errule{lift$_\Xi$} and \errule{$\Xi$} implement the overloading of $\Pi$ and $\Sigma$. The remaining rules (relegated to \Appendix{sec:extra-typing-rules}) implement congruences for the type constructors and ensure that $\TEqvJ \Delta \tau \tau \kappa$ is an equivalence relation.

\subsubsection*{Typing}

\begin{figure}
\begin{small}
\begin{gather*}
\fbox{$\TypeJ \Delta \Phi \Gamma M \tau$}
\\
\ib{\irule[\trule{var}]
          {\PEnvJ \Delta \Phi}
          {\TEnvJ \Delta \Gamma}
          {x : \tau \in \Gamma};
          {\TypeJ \Delta \Phi \Gamma x \tau}}
\isp
\ib{\irule[\trule{k}]
          {\PEnvJ \Delta \Phi}
          {\TEnvJ \Delta \Gamma}
          {K : \sigma};
          {\TypeJ \Delta \Phi \Gamma K \sigma}}
\isp
\ib{\irule[\trule{conv}]
          {\TypeJ \Delta \Phi \Gamma M \tau}
          {\TEqvJ \Delta \tau \upsilon \TypeK};
          {\TypeJ \Delta \Phi \Gamma M \upsilon}}
\\
\ib{\irule[\trule{$\I\to$}]
          {\KindJ \Delta {\tau} {\TypeK}}
          {\TypeJ \Delta \Phi {\Gamma, x : \tau} M {\upsilon}};
          {\TypeJ \Delta \Phi \Gamma {\lambda x \co \tau. M} {\tau \to \upsilon}}}
\rsp
\ib{\irule[\trule{$\E\to$}]
          {\TypeJ \Delta \Phi \Gamma M {\tau \to \upsilon}}
          {\TypeJ \Delta \Phi \Gamma N {\tau}};
          {\TypeJ \Delta \Phi \Gamma {M \, N} {\upsilon}}}
\\
\ib{\irule[\trule{$\I\then$}]
          {\PredJ \Delta \pi}
          {\TypeJ \Delta {\Phi, \pi} \Gamma M \tau};
          {\TypeJ \Delta \Phi \Gamma {\Lambda v \co \pi. M} {\pi \then \tau}}}
\rsp
\ib{\irule[\trule{$\E\then$}]
          {\TypeJ \Delta \Phi \Gamma M {\pi \then \tau}}
          {\EntJ \Delta \Phi Q \pi};
          {\TypeJ \Delta \Phi \Gamma {\AppT M Q} \tau}}
\\
\ib{\irule[\trule{$\I\forall$}]
          {\TypeJ {\Delta, \alpha : \kappa} \Phi \Gamma M \tau};
          {\TypeJ \Delta \Phi \Gamma {\Lambda \alpha \co \kappa. M} {\forall \alpha\co\kappa. \tau}}}
\rsp
\ib{\irule[\trule{$\E\forall$}]
          {\TypeJ \Delta \Phi \Gamma M {\forall \alpha\co\kappa. \tau}}
          {\KindJ \Delta \upsilon \kappa};
          {\TypeJ \Delta \Phi \Gamma {\AppT M \upsilon} {\tau[\upsilon/\alpha]}}}
\rsp
\ib{\irule[\trule{$\I\#$}]
         {\KindJ \Delta \zeta \kappa};
         {\TypeJ \Delta \Phi \Gamma {\Sing \zeta} {\Sing \zeta}}}
\\
\ib{\irule[\trule{\I{\Xi\,}}]
          {\begin{array}{@{}c@{}}
             {\KindJ \Delta \xi \LabK} \isp
             {\TypeJ \Delta \Phi \Gamma N \tau} \\
             {\TypeJ \Delta \Phi \Gamma M {\Sing\xi}}
           \end{array}};
          {\TypeJ \Delta \Phi \Gamma {\LabTermX \Xi M N} {\Xi \Row {\LabTy \xi \tau}}}}
\rsp
\ib{\irule[\trule{\E{\Xi\,}}]
          {\begin{array}{@{}c@{}}
             {\TypeJ \Delta \Phi \Gamma M {\Xi {\Row {\LabTy \xi \tau}}}} \\
             {\TypeJ \Delta \Phi \Gamma N {\Sing \xi}}
           \end{array}};
          {\TypeJ \Delta \Phi \Gamma {\UnlabelX \Xi M N} \tau}}
\qquad
(\Xi \in \Set{\Pi, \Sigma})
\\[5pt]
\fbox{$K : \sigma$}
\\
\begin{aligned}
  \Prj{} &: \forall y z \co \RowK \TypeK. \, \LeqP y z \then \Pi z \to \Pi y \\
  (\Concat) &: \forall x y z \co \RowK \TypeK. \, \PlusP x y z \then \Pi x \to \Pi y \to \Pi z \\
  \Inj &: \forall y z \co \RowK \TypeK. \, \LeqP y z \then \Sigma y \to \Sigma z \\
\end{aligned}
\qquad\qquad\qquad
\begin{aligned}
  \In &: \forall f \co {\TypeK \to \TypeK}. f \, (\mu f) \to \mu f \\
  \Out &: \forall f \co {\TypeK \to \TypeK}. \mu f \to (f \, (\mu f)) \\
  \Fix &: \forall t \co \TypeK. (t \to t) \to t
\end{aligned}
\\
\begin{aligned}
  (\Branch) &: \forall x y z \co \RowK \TypeK, t \co \TypeK. \, \PlusP x y z \then (\Sigma x \to t) \to (\Sigma y \to t) \to (\Sigma z \to t) \\
  \mathsf{syn}^{(\kappa)} &: \forall f \co \kappa \to \TypeK, z \co \RowK \kappa. \Sing f \to (\forall l \co \LabK, t \co \kappa. \LeqP {\Row {\LabTy l t}} z \then \Sing l \to f \, t) \to \Pi (f \, z) \\
  \mathsf{ana}^{(\kappa)} &: \forall f \co \kappa \to \TypeK, z \co \RowK \kappa, t \co \TypeK. \Sing f \to (\forall l \co \LabK, u \co \kappa. \LeqP {\Row {\LabTy l u}} z \then \Sing l \to f \, u \to t) \to \Sigma (f \, z) \to t
\end{aligned}
\end{gather*}
\end{small}
\caption{Typing}
\label{fig:typing}
\end{figure}

Typing rules are given in \cref{fig:typing}. Much of the interest is deferred to the constants, discussed next. We make type abstraction and application and predicate abstraction and application explicit here. We assume that type abstraction and application would typically, and predicate abstraction and application would always, be implicit in a surface language. $(\LabTermX \Pi - -)$ and $(\UnlabelX \Pi - -)$ construct and deconstruct singleton records, and $(\LabTermX \Sigma - -)$ and $(\UnlabelX \Sigma - -)$ construct and deconstruct singleton variants. These essentially provide the ``base cases'' of our record and variant system; we will construct larger records, or eliminate larger variants, by combining these individual cases.

\subsubsection*{Constants}

The types of $\In$, $\Out$, and $\Fix$ are standard. The $\Prj$ operator projects smaller records from largers ones. This operator is overloaded in both its input and output types. Given $M : \Pi z$, a record built following row $z$, $\Prj\,M$ can build \emph{any} subrecord of $M$, from the empty record to the full subrecord $M$ itself. The usual record selection operator is a special case of our projection operator, and is definable in \Rome~\cref{sec:first-class-labels}. The $(\Concat)$ operator concatenates records, so long as their types fit together. The particulars of the predicates $\LeqP {y} {z}$ and $\PlusP {x} {y} {z}$ will be discussed shortly~\cref{sec:entailment}. The dual operators for variants are $\Inj$ and $(\Branch)$. The $\Inj$ operator injects smaller variants into larger variants; as with $\Prj$, it is overloaded in both its input and output types. The $(\Branch)$ operator combines variant eliminators: if $f$ eliminates variants over row $x$, and $g$ eliminates variants over $y$, then $f \Branch g$ (read ``$f$ or $g$'') eliminates variants built from the combination of $x$ and $y$. Finally, $\Syn$ and $\Ana$ provide ``label-generic'' construction of record and elimination of variants. Intuitively, $\Syn \, \Sing f \, M$ builds a record where each field can be built by $M$, and $\Ana \, \Sing f \, M$ eliminates a variant where each case can be handled by $M$. We will describe these operators in more detail later in this section~\cref{sec:label-generic-programming}.

\subsection{Entailment}
\label{sec:entailment}

We describe \Rome's predicates and entailment relation. Predicates play a central role in understanding rows in \Rome. In types, they capture the structure and manipulation of rows, via containment ($\lesssim$) and combination ($\odot$) relations. In terms, the proofs of these relations determine the meaning of the projection, concatenation, injection, and branching primitives. Proofs of the entailment relation are represented in terms by \emph{evidence}.

\begin{figure}
\begin{gather*}
\begin{array}{r@{\hspace{5px}}l@{\qquad}r@{\hspace{5px}}l@{\qquad}r@{\hspace{5px}}l@{\qquad}r@{\hspace{5px}}l}
  \text{Evidence variables} & v \in \mathcal V &
  \text{Indices} & i, j, k, m, n \in \mathbb N &
  \text{Index maps} & p, q \in \mathbb N \pto \mathbb N
\end{array}
\\[5pt]
\begin{doublesyntaxarray}
  \text{Evidence} & \mcr{\mathcal Q \ni  Q} & ::= & v \mid \TransV Q Q \mid \InclV \, p \mid \CombV \, p \, q \\
  & & & \mid & \LeqV {refl} \mid \LeqV {map} \, Q \mid \LeqV{plusL} \, Q \mid \LeqV {plusR} \, Q \\
  & & & \mid & \PlusV {emptyL} \mid \PlusV {emptyR} \mid \PlusV {map} \, Q \mid \PlusV {complL} \, Q \mid \PlusV {complR} \, Q
\end{doublesyntaxarray}
\end{gather*}
\caption{Syntax: evidence}
\label{fig:syntax-evid}
\end{figure}

The syntax of evidence is given in \cref{fig:syntax-evid}. The essential data carried by evidence are index maps, mapping entries in one row to entries in another. We will write index maps using $\mapsto$, to avoid confusion between semantic objects and syntactic $\lambda$ definitions, and will write $\ran(p)$ for the range of an index map. In practice, index maps could be implemented as vectors of indices, rather than as functions. As we insist that rows are ordered, index maps will always be monotononic. The forms of evidence otherwise correspond directly to the forms of entailment proof.

\begin{figure}
\begin{small}
\begin{gather*}
\fbox{$p : \rho \subseteq \rho$}
\\
\ib{\irule{\xi_k = \xi'_{p(k)}}
          {\tau_k = \tau'_{p(k)}};
          {p : \RowIx i 0 m {\LabTy {\xi_i} {\tau_i}} \subseteq \RowIx j 0 n {\LabTy {\xi'_j} {\tau'_j}}}!
          {\text{for all $k \in 0 \dots m$}}}
\\[5pt]
\fbox{$\EntJ \Delta \Phi Q \pi$}
\\
\ib{\irule[\entrule{ax}]
          {v\co\pi \in \Phi};
          {\EntJ \Delta \Phi v \pi}}
\rsp
\ib{\irule[\entrule{refl}]
          {\PEnvJ \Delta \Phi};
          {\EntJ \Delta \Phi {\LeqV {refl}} {\LeqP \rho \rho}}}
\rsp
\ib{\irule[\entrule{conv}]
          {\EntJ \Delta \Phi Q {\pi_1}}
          {\PEqvJ \Delta {\pi_1} {\pi_2}};
          {\EntJ \Delta \Phi Q {\pi_2}}}
\\
\ib{\irule[\entrule{trans}]
          {\EntJ \Delta \Phi {Q_1} {\LeqP {\rho_1} {\rho_2}}}
          {\EntJ \Delta \Phi {Q_2} {\LeqP {\rho_2} {\rho_3}}};
          {\EntJ \Delta \Phi {\TransV {Q_1} {Q_2}}{\LeqP {\rho_1} {\rho_3}}}}
\rsp
\ib{\irule[\entsrule{$\odot\Compl\Ri$}]
          {\EntJ \Delta \Phi Q {\LeqP {\rho_1} {\rho_2}}};
          {\EntJ \Delta \Phi {\PlusV {complR} \, Q} {\PlusP {\rho_1} {\rho_2 \Compl \rho_1} {\rho_2}}}}
\\
\ib{\irule[\entsrule{$\odot\Compl\Le$}]
          {\EntJ \Delta \Phi Q {\LeqP {\rho_1} {\rho_2}}};
          {\EntJ \Delta \Phi {\PlusV {complL} \, Q} {\PlusP {\rho_2 \Compl \rho_1} {\rho_1} {\rho_2}}}}
\rsp
\ib{\irule[\entsrule {$\lesssim$}]
          {p : {\RowIx i 0 m {\LabTy {\xi\One_i} {\tau\One_i}}} \subseteq {\RowIx j 0 n {\LabTy {\xi\Two_j} {\tau\Two_j}}}};
          {\EntJ \Delta \Phi {\InclV \, p} {\LeqP {\RowIx i 0 m {\LabTy {\xi\One_i} {\tau\One_i}}} {\RowIx j 0 n {\LabTy {\xi\Two_j} {\tau\Two_j}}}}}}
\\
\ib{\irule[\entsrule {$\odot$}]
          {p : {\RowIx i 0 m {\LabTy {\xi\One_i} {\tau\One_i}}} \subseteq {\RowIx k 0 n {\LabTy {\xi\Three_k} {\tau\Three_k}}}}
          {q : {\RowIx j 0 {n - m} {\LabTy {\xi\Two_j} {\tau\Two_j}}} \subseteq {\RowIx k 0 n {\LabTy {\xi\Three_k} {\tau\Three_k}}}};
          {\EntJ \Delta \Phi {\CombV \, p \, q}
            {\PlusP
                {\RowIx i 0 m {\LabTy {\xi\One_i} {\tau\One_i}}}
                {\RowIx j 0 {n - m} {\LabTy {\xi\Two_j} {\tau\Two_j}}}
                {\RowIx k 0 n {\LabTy {\xi\Three_k} {\tau\Three_k}}}}}!
          {\ran(p) \cap \ran(q) = \emptyset}}
\\
\ib{\irule[\entrrule{$\EmptyRow_{\odot\mathsf L}$}]
          {\PEnvJ \Delta \Phi};
          {\EntJ \Delta \Phi {\PlusV {emptyL}} {\PlusP {\EmptyRow} {\rho} {\rho}}}}
\rsp
\ib{\irule[\entrrule{$\EmptyRow_{\odot\mathsf R}$}]
          {\PEnvJ \Delta \Phi};
          {\EntJ \Delta \Phi {\PlusV {emptyR}} {\PlusP \rho \EmptyRow \rho}}}
\\
\ib{\irule[\entrrule{$\lesssim$map}]
          {\EntJ \Delta \Phi Q {\LeqP {\rho_1} {\rho_2}}};
          {\EntJ \Delta \Phi {\LeqV {map} \, Q}{\LeqP {\Map\phi\,\rho_1} {\Map\phi\,\rho_2}}}}
\rsp
\ib{\irule[\entrrule{$\odot$map}]
          {\EntJ \Delta \Phi Q {\PlusP {\rho_1} {\rho_2} {\rho_3}}};
          {\EntJ \Delta \Phi {\PlusV {map} \, Q} {\PlusP {\Map\phi\,\rho_1} {\Map\phi\,\rho_2} {\Map\phi\,\rho_3}}}}
\\
\ib{\irule[\entsrule{${\odot}{\lesssim}\Le$}]
          {\EntJ \Delta \Phi Q {\PlusP {\rho_1} {\rho_2} {\rho_3}}};
          {\EntJ \Delta \Phi {\LeqV {plusL} \, Q} {\LeqP {\rho_1} {\rho_3}}}}
\rsp
\ib{\irule[\entsrule{${\odot}{\lesssim}\Ri$}]
          {\EntJ \Delta \Phi Q {\PlusP {\rho_1} {\rho_2} {\rho_3}}};
          {\EntJ \Delta \Phi {\LeqV {plusR} \, Q} {\LeqP {\rho_2} {\rho_3}}}}
\end{gather*}
\end{small}
\caption{Entailment}
\label{fig:entailment}
\end{figure}

The entailment relation is given in \cref{fig:entailment}. We use a witnessed containment judgment $p : \rho_1 \subseteq \rho_2$, denoting that index map $p$ locates the entries of $\rho_1$ in $\rho_2$. The judgment $\EntJ \Delta \Phi Q \pi$ holds when $Q$ is evidence for $\pi$ given assumptions in $\Phi$. The base cases of entailment are for row literals. Rule \entrule{$\lesssim$} states that one row is contained within another exactly when there is a map from the entries of the contained row to entries of the containing row. Rule \entrule{$\odot$} states that $\rho_3$ is the combination of $\rho_1$ and $\rho_2$ when the entries of each of $\rho_1$ and $\rho_2$ can be mapped to disjoint entries of $\rho_3$. Rules (\textsc{n-$\EmptyRow_{\odot\mathsf L}$},\textsc{n-$\EmptyRow_{\odot\mathsf R}$}) make the empty row the unit of combination. These rules are derivable for concrete $\rho$, and they allow us to derive $\LeqP \EmptyRow \rho$ for any $\rho$. Rules (\textsc{n-$\odot\lesssim\Le$},\textsc{n-$\odot\lesssim\Ri$}) describe one direction of the relationship between containment and combination: if $\rho_1$ and $\rho_2$ combine to make $\rho_3$, then each of $\rho_1$ and $\rho_2$ must be contained in $\rho_3$. Rules (\textsc{n-$\odot\Compl\Le$},\textsc{n-$\odot\Compl\Ri$}) give the other direction: if $\rho_1$ is contained in $\rho_2$, then there is some row which can be combined with $\rho_1$ to make $\rho_2$. The remaining rules are unsurprising.

Many of the expected properties of the entailment relation, such as reflexivity and transitivity of containment, hold by construction. Others holds by simple inductive arguments. In particular: entailment is preserved under substitution, and the $\odot$ relation is commutative.

\subsection{Variants and Records}
\label{sec:variants-and-records}
\InlineOn

\Rome's primitive operations are quite general, and it may not be immediately apparent how they would be used in practice. We will now walk though several relatively simple examples of programming with \Rome. The code shown in these sections is as type checked and executed by our experimental implementation of \Rome\ifminted.\else, modulo some substitution of mathematical symbols (such as $\forall$ and $\Sigma$) for keywords (such as \texttt{forall} and \texttt{Sigma}).\fi

The starting point for work on the \textsc{Rose} type system was Wand's~\citep{Wand87} encoding of multiple inheritance using row types. The essence of Wand's example is expressed in \Rome as:
\begin{rosi}
wand : forall x y z t. x + y ~ z, {'l := t} < z => Pi x -> Pi y -> t
wand = \ r s. prj (r ++ s) / #'l
\end{rosi}
The term !wand! is a function that takes two record arguments !m! and !n!. It concatenates these records, and projects the field labeled !'l! from the result. In our concrete syntax, we use initial quote marks to distinguish label constants from label variables. The surrounding singleton elimination form !/ #'l! determines the output type of the !prj! operator. (In our implementation, we overload the !/! and !:=! symbols to apply to either variants or records.) This operation makes sense if either !m! or !n! contains an !'l!-labeled field. This is captured in the type of !wand!: !wand! takes two records, one of type !Pi x! and one of type !Pi y! and return a value of type !t!, \emph{so long as} (1) the rows !x! and !y! can be combined to make a row !z!, and (2) the row !z! contains an entry with label !'l!. Crucially, we do not constrain which of !x! and !y! originally contained label !'l!.

Records and variants are dual, and so there is a dual version of Wand's problem in terms of variants. It is expressed in \Rome as:
\begin{rosi}
dnaw : forall x y z t u. x + y ~ z, {'l := t} < z => (Sigma x -> u) -> (Sigma y -> u) -> t -> u
dnaw = \ f g x. (f | g) (inj (#'l := x))
\end{rosi}
The term !dnaw! is a function of two variant eliminators !f! and !g! and a value !x!. It combines these eliminators, and then passes to the combined eliminator a variant with constructor !'l! and value !x!. This operation makes sense so long as either !f! or !g! includes constructor !'l! among the cases it handles. This is captured in its type: !dnaw! takes two eliminators, one from !Sigma x! to !u! and one from !Sigma y to u!, and returns a !t -> u! function, \emph{so long as} the rows !x! and !y! can be combined to make !z!, and !z! contains an entry !l := t!.

\subsection{First-Class Labels}
\label{sec:first-class-labels}

Reusable coding patterns, such as selecting a single field from a record or constructing a variant from a given label, are already apparent in our simple \Rome examples. \citet{HubersM23} extended \textsc{Rose} with support for first-class labels, allowing these patterns to be captured in general. \cref{fig:fcl} shows some useful utility functions with first-class labels.

\begin{figure}
\begin{minipage}[t]{0.4\linewidth}
\begin{rosi}
sel : forall l t z. {l := t} < z =>
      Pi z -> #l -> t
sel = \ r l. prj r / l
\end{rosi}
\codesep
\begin{rosi}
con : forall l t z. {l := t} < z =>
      #l -> t -> Sigma z
con = \ l x. inj (l := x)
\end{rosi}
\end{minipage}
\begin{minipage}[t]{0.59\linewidth}
\begin{rosi}
rcon : forall l f z. {l := f} < z =>
       #l -> f (Mu (Sigma z)) -> Mu (Sigma z)
rcon = \ l x. in (con l x)
\end{rosi}
\codesep
\begin{rosi}
case : forall l t u.
       #l -> (t -> u) -> Sigma {l := t} -> u
case = \ l f x. f (x / l)
\end{rosi}
\codesep
\begin{rosi}
match : forall t u. t -> (t -> u) -> u
match = \ x f. f x
\end{rosi}
\end{minipage}
\caption{Common operations with first-class labels}
\label{fig:fcl}
\end{figure}

The !sel! and !con! functions abstract field selection and variant construction. We use singleton types !#l! to make the otherwise-implicit label explicit. In this paper, we will be interested in recursive variants, so we also define a recursive variant construction function !rcon!.

The !case! function abstracts a common pattern in defining variant eliminators. For example, define a type of Booleans as so:
\begin{rosi}
type Bool : *
type Bool = Sigma { 'True := Unit, 'False := Unit }
\end{rosi}
(Definitions for the !Unit!, !Nat!, and !List! types can be found in \Appendix{sec:extra-data-types}.) We write the !not! function by combining individual eliminators for the singleton variants !Sigma! !{ 'True! !:=! !Unit }! and !Sigma { 'False := Unit }!:
\begin{rosi}
not : Bool -> Bool
not = case #'True (con #'False)
    | case #'False (con #'True)
\end{rosi}
This style is quite pleasant when programming in a points-free style. However, it can become cumbersome to read when the scrutinee needs to be provided explicitly. The !match! function (actually just reverse application) makes the code much easier to read:
\\
\begin{minipage}{\linewidth} %
\begin{rosi}
not = \ b. match b
           ( case #'True (con #'False)
           | case #'False (con #'True))
\end{rosi}
\end{minipage}

The appeal of the fully generic !prj! and !inj! functions may not be apparent, given that we have immediately wrapped them to provide more familiar field selection and variant construction operations. Consider the following function, which applies a modifier function to one field of a record:
\begin{rosi}
modify : forall l t u y z1 z2. {'l := t} + y ~ z1, {'l := u} + y ~ z2 =>
         #l -> (t -> u) -> Pi z1 -> Pi z2
modify = \ l f r. (l := f (sel l r)) ++ prj r
\end{rosi}
This definition makes two uses of !prj!: in !sel! !l! !r!, !prj! is used to select a single record field; the other use of !prj! computes the remainder of the record less field !l!. In many row type systems, these would have to be distinct primitive operations. For us, they are just two uses of the same underlying operator.

\subsection{Label-Generic Programming}
\label{sec:label-generic-programming}

The operations we have discussed so far all operate on specific labels. \citet{HubersM23} generalize \textsc{Rose} with two label-generic operators. The \emph{synthesis} operator generates a record with arbitrary fields, given a term that can construct a value for any field. The \emph{analysis} operator eliminates a variant with arbitrary constructors, given a term that can eliminate a value for any constructor. We will illustrate the use of these operators using our encoding of functors in \Rome.

Many encodings of extensible recursive data types rely on building data types from functors, and ``tying the knot'' of the recursive data type at the last possible moment. \citet{Swierstra08} makes this pattern explicit, introducing a (binary) coproduct operator $\oplus$ to combine functors, and an instance that concludes $\mathsf{Functor}\,(f \oplus g)$ from instances for $\mathsf{Functor} \, f$ and $\mathsf{Functor} \, g$. In turn, the functor structure is used to implement various traversals and eliminations of recursive variants.

We would like to apply a similar approach in \Rome, but with rows of functors rather than binary combinations. As with Swierstra's coproduct, the key feature will be a witness that, if $\rho$ is a row of functors, then $\Sigma \rho$ (and $\Pi \rho$) are functorial as well. The challenge is that, while we need to impose some structure on the types in $\rho$---namely, that they are functorial---we do not want to limit the labels in $\rho$.

First, we define functoriality. The type !Functor f! is the the type of the mapping operator for a type constructor !f!:
\begin{rosi}
type Functor : (* -> *) -> *
type Functor = \f. forall a b. (a -> b) -> f a -> f b
\end{rosi}
Next, we define the mapping operator !fmapS! for variants built from functors:
\begin{rosi}
fmapS : forall z : R[* -> *]. Pi (Functor z) -> Functor (Sigma z)
fmapS = \ d. /\ a b. \ f w.
        ana #(\x. x a) (\ l x. con l (sel d l f x)) w
\end{rosi}
Our goal is a value of type !Functor (Sigma z)!, that is to say, one of type !forall a b. (a -> b) -> (Sigma z) a! !->! !(Sigma z) b!. We are relying on our overloading of !Sigma! to apply at higher kinds here: !(Sigma z) a! is the variant built from the application of each type operator in !z! to argument !a!. Obviously, we need to know that all the types in !z! are functorial themselves. We capture this requirement as the argument !Pi (Functor z)!. We (implicitly) map type operator !Functor! over row !z!. The resulting row contains the type of the !fmap! operator for each type constructor in !z!. We then construct a record of these !fmap!s.

Our implementation uses the label-generic variant eliminator !ana!. We focus on its second argument, which is the ``body'' of the eliminator. The type of the body is highly polymorphic; in this instance, it has type !forall l u. {l := u} < z => #l -> u a -> Sigma (z b)!. That is: for an arbitrary !l := u! in the input row !z!, we must transform a !u a! value to a !Sigma (z b)! result. To do so, we need a mapping operator for the type !u!; luckily, the record of mapping operators !d! must contain just such an operator, in field !l!. The expression !sel d l f x! selects this mapping operator from !d!, and then applies it to !f! and !x!, giving a term of type !u b!. To get a value of type !Sigma (z b)!, we need only reapply the constructor !l!.

There is one significant unexplained step in this account. The result type !Sigma (z b)! comes from the expected type of !fmapS! itself. However, the input is of type !(Sigma z) a!; how did we get a constraint !{l := u} < z! instead of something in terms of $\mathtt z^\$ \, \mathtt a$ (as rule \errule{lift$_\Sigma$} would suggest)? The answer is the provided type operator !\ x. x a!, which is applied to the type of the analyzed row. This operator aligns the type we need in the body (!u a!) with the constraint we would like (in terms of just !u!).

The case for products is dual:
\begin{rosi}
fmapP : forall z : R[* -> *]. Pi (Functor z) -> Functor (Pi z)
fmapP = \ d. /\ a b. \f r.
        syn #(\x. x b) (\l. sel d l f (sel r l))
\end{rosi}
The type is constructed similarly, and the same problem arises: How are we to construct a record of type !Pi (z b)! when we know nothing about the labels in !z!? The answer is the label-generic record constructor !syn!. We focus on its second argument. The type is similarly generic: !forall l u. {l := u} < z => #l -> u b!. To construct the needed result, we begin by projecting a value of type !u a! from the input record !r!, and then using the !l!-labeled mapping function to transform it to a value of type !u b!. Again, the provided type operator !\ x. x b! aligns the required result type !u b! with the needed constraint !{l := u} < z!.

\subsection{Row Complements}
\label{sec:row-complements}

This work extends the existing \RO type system with first-class row complements. To see why, we consider another example of label-generic programming.The goal is generic construction of equality functions for variant types. The type !Eq t! describes an equality function for type !t!:
\begin{rosi}
type Eq : * -> *
type Eq = \ t. t -> t -> Bool
\end{rosi}
The !eqS! function compares arbitrary variants !Sigma z!, so long as we have comparison functions for each entry in !z!:
\begin{rosi}
eqS : forall z : R[*]. Pi (Eq z) -> Eq (Sigma z)
eqS = \ d v w. ana #(\ x. x)
                   (\ l y. match v
                           ( case l (\ x. sel d l x y)
                           | const False)) w
\end{rosi}
The logic, captured in the second argument to !ana!, is as follows. We are given the constructor !l! and contents !y! of the variant !w!. We then branch on !v!. If !v! has constructor !l! and body !x!, we must then compare !x! and !y!. We do so using the !l!-labeled equality function in !d!. Otherwise, !v! and !w! were built using different constructors, and so must be unequal.

The challenge here arises in typing the default branch !const False!. All we know about the type !Sigma z! of !w! and !v! is that there is some !t! such that !{l := t} < z!. In particular, we do not have a name for the ``rest'' of !z!. However, !const False! needs to have type !Sigma y -> Bool!, where !y! is exactly the ``rest'' of !z!.

\RO addressed this gap by complicating the type of !ana! (and !syn!), replacing the !x < z! constraint with a constraint !x + y ~ z! for a new type parameter !y!. In addition to being cumbersome, this solution does not scale. For example, knowing that !x + y ~ z! and that !z + w ~ v! does not provide a solution to !x + ? ~ v!, even though such a row must exist.

Our solution is to make row complements first class: $\rho_{2} \Compl \rho_{1}$ contains the entries of $\rho_{2}$ that are not in $\rho_{1}$. Of course, without knowing anything about the relationship between $\rho_{2}$ and $\rho_{1}$, it is difficult to say anything about $\rho_{2} \Compl \rho_{1}$. However, if we know that $\rho_{1} \lesssim \rho_{2}$, then we can do better. In particular, rule \entsrule{$\odot\Compl\Ri$} allows us to conclude that $\rho_{1} \odot (\rho_{2} \Compl \rho_{1}) \sim \rho_{2}$.

This addresses the problem in the typing of !eqS!: we can now give !const False! the type !Sigma (z - {l! !:=! !t}) -> Bool!. It also addresses the scaling problem that came from replacing !x < z! predicates with !x + y ~ z!. On the one hand, given !x < y! and !y < z!, we can conclude !x < z!; given the latter, we also have !x + (z - x) ~ z!.

It may be surprising that we allow $\rho_{2} \Compl \rho_{1}$ at all without knowing that $\rho_{1} \lesssim \rho_{2}$. Our reasons for doing so are several. Attempting to impose such a restriction would make the definitions of kinding and predicate formation mutually recursive. Moreover, predicates have evidence; should the evidence used to justify $\rho_{2} \Compl \rho_{1}$ be reflected somehow in the type? Do different justifications of $\rho_{1} \lesssim \rho_{2}$ somehow distinguish instances of $\rho_{2} \Compl \rho_{1}$? Allowing $\rho_{2} \Compl \rho_{1}$ regardless of the relationship between $\rho_{2}$ and $\rho_{1}$ avoids all these problems, and poses no difficult in the definition of the complement.

\InlineOff
\section{Extensible Recursive Functions}
\label{sec:xr}
\InlineOn

\subsection{The Extensible Recursion Problem}

Invoking \citet{Wadler98}, we give the following definition of an extended expression problem:

\begin{quote}
  The Extended Expression Problem is a new name for an old problem.  The goal is to define datatypes by cases, where one can combine datatype and combine functions over the datatypes, without recompiling existing code, and while retaining static type safety.  For the concrete example, we take expressions as the data type, begin with one case (arithmetic terms) and one function (evaluation), then add two more cases (Boolean and functional terms) and one more function (desugaring Boolean terms).
\end{quote}

\noindent
This section describes our solution to the extended expression problem. We begin with an overview of the technical challenges in solving the extended expression problems, introduce extensible histomorphisms, a new abstraction of recursive traversals that supports extensibility, and then demonstrate our implementations of evaluation and desugaring.

We structure expressions by:\\[-6pt]
\begin{minipage}[t]{0.51\linewidth}
\begin{rosi}
type ArithF : R[* -> *]
type ArithF = { 'IConst := Const Nat
              , 'Plus := (\ t. Pair t t) }
\end{rosi}
\codesep
\begin{rosi}
type BoolF : R[* -> *]
type BoolF = { 'BConst := Const Bool
             , 'If := (\ t. Triple t t t) }
\end{rosi}
\end{minipage}
\begin{minipage}[t]{0.47\linewidth}
\begin{rosi}
type LamF : R[* -> *]
type LamF = { 'Var := Const Nat
            , 'Lam := Id
            , 'App := (\t. Pair t t) }
\end{rosi}
\end{minipage}
\\
Arithmetic expressions include constants and sums. Boolean expressions include constants and conditionals. The !Pair! type is defined as !\ a b. Pi { '1 := a, '2 := b }!, and similarly for !Triple!. Functional expressions include variables, abstractions, and applications. We refer to variables by de Bruijn indices; consequently, abstractions need to carry no information other than their body. For example, the negation function can be given by:
\begin{rosi}
notE : forall z. LamF < z, BoolF < z => Mu (Sigma z)
notE = rcon #'Lam (rcon #'If (triple (rcon #'Var zero)
                                     (rcon #'BConst False)
                                     (rcon #'BConst True)))
\end{rosi}
By giving !notE! a generic type, we avoid the need to manually inject it into larger expression types.

To illustrated the difficulties in the extended expression problem, consider na\"ive evaluation functions for arithmetic expressions.
\begin{rosi}
naiveA : Mu (Sigma ArithF) -> Nat
naiveA = fix (\ rec exp. match (out exp)
  ( case #'IConst id
  | case #'Plus (\ p. add (rec (sel p #'1)) (rec (sel p #'2)))))
\end{rosi}
and Boolean expressions:
\begin{rosi}
naiveB : Mu (Sigma BoolF) -> Bool
naiveB = fix (\ rec exp. match (out exp)
  ( case #'BConst id
  | case #'If (\ t. match (rec (sel t #'1))
      ( case #'True (const (rec (sel t #'2)))
      | case #'False (const (rec (sel t #'3)))))))
\end{rosi}
To be clear: these are well-typed definitions of big-step evaluation functions for the arithmetic and Boolean langauges. We list the reasons it is not immediately possible to write something like !naiveA | naiveB! as an evaluator for the combined arithmetic and Boolean langauges:

First, the output types do not align. We could (and will) address this by introducing a variant for values. In doing so, we will also have to account for recursive calls returning the wrong type of value. This adds some incidental complexity, but does not change anything fundamental.

\newcommand\ArithF{\mathtt{ArithF}}
\newcommand\BoolF{\mathtt{BoolF}}
\newcommand\ABF{\mathtt{ABF}}

More significantly, the input types do not align. The outer !Mu! wrapper is not the problem; we could easily adapt !naiveA! to be of type !Sigma ArithF (Mu (Sigma ArithF)) -> Nat!, and similarly for !naiveB!. The problem is in the recursive  instances. Writing !ABF! for the combination of !ArithF! and !BoolF!, we have that
\(
  \PlusP {\Sigma \, \ArithF \, (\mu \, (\Sigma \, \ABF))} {\Sigma \, \BoolF \, (\mu \, (\Sigma \, \ABF))} {\Sigma \, \ABF {(\mu \, (\Sigma \, \ABF))}},
\)
but not that
\(
  \PlusP {\Sigma \, \ArithF \, (\mu \, (\Sigma \, \ArithF))} {\Sigma \, \BoolF \, (\mu \, (\Sigma \, \BoolF))} {\Sigma \, \ABF {(\mu \, (\Sigma \, \ABF))}},
\)
as we would need to combine !naiveA! and !naiveB!. In essence, the problem is the recursive calls. The recursive calls in !naiveA! are to !naiveA!, not to the combined !naiveA! and !naiveB!. However, we cannot be sure that the expressions inside an arithmetic expression are only arithmetic expressions; indeed, the !notE! example interleaves terms from the functional and Boolean languages.

At this point, it might be tempting to suggest a branching operator that can ``unfold'' fixpoints, and refold them with a more general type for recursive calls. At the end of the section, we will demonstrate that, even with such a capability, we could not generally combine eliminations of recursive variants.

\subsection{Algebras and Extensibility}

Our approach to describing extensible recursive functions is to add a layer of abstraction, following the well-studied idea of recursion schemes \citep{MeijerFP91,Mendler91}. The simplest recursion scheme is the catamorphism, whose \Rome description is given by:\\[-6pt]
\begin{minipage}[t]{0.4\linewidth}
\begin{rosi}
type FAlg : (* -> *) -> * -> *
type FAlg = \ f a. f a -> a
\end{rosi}
\end{minipage}
\begin{minipage}[t]{0.59\linewidth}
\begin{rosi}
cata : forall f a. Functor f -> FAlg f a -> Mu f -> a
cata = \ map a. fix (\ cata x. a (map cata (out x)))
\end{rosi}
\end{minipage}
Catamorphisms are uniquely determined by $F$-algebras, functions of type $f \, a \to a$, given a type family $f$ and result type $a$. Intuitively, this consists of one step of a bottom-up fold. A catamorphism with type !Mu f -> a! is built from an $F$-algebra and a mapping function !Functor f! to access the data.

$F$-algebras are innately well-suited to extensibility. An $F$-algebra for computing the size (i.e., number of constructors) of arithmetic terms is:
\begin{rosi}
sizeA : FAlg (Sigma ArithF) Nat
sizeA = case #'IConst (\u. one)
       | case #'Plus (\t. succ (add (sel t #'1) (sel t #'2)))
\end{rosi}
and one for Boolean terms is:
\begin{rosi}
sizeB : FAlg (Sigma BoolF) Nat
sizeB = case #'BConst (\u. one)
      | case #'If (\t. succ (add (sel t #'1) (add (sel t #'2) (sel t #'3))))
\end{rosi}
These are individually straightforward. Expanding their type, we have that !sizeB! has type !Sigma BoolF! !Nat! !-> Nat!, !sizeA! has type !Sigma ArithF -> Nat!, and so !sizeA | sizeB! can be given type !Sigma ABF Nat ->! !Nat!, just as we would hope.

Unfortunately, not all recursive functions are catamorphic. Evaluating arithmetic expressions is naturally catamorphic. Evaluating Boolean expressions can be seen catamorphically, even if it may seem odd to preemptively evaluate both branches of the conditional. Evaluating functions, however, cannot be viewed catamorphically: evaluating an abstraction should not be defined in terms of evaluating its body. To capture functions like big-step evaluation, we need a more expressive recursion scheme.

Our proposed recursion scheme, which we call an extensible histomorphism, is defined by:
\begin{rosi}
type Xh : R[* -> *] -> (R[* -> *] -> *) -> *
type Xh = \ z f. forall w. z < w => Sigma z (Mu (Sigma w)) -> (Mu (Sigma w) -> f w) -> f w
\end{rosi}
Our most immediate inspiration are Mendler histomorphisms~\citep{Mendler91}, although our final scheme is some distance from Mendler's. An extensible histomorphic algebra !Xh z f! (henceforth, simply !Xh! \emph{algebra}) is defined in terms of a row of type constructors !z!, the cases of the variant to be traversed, and an output type function !f!. These pieces are tied together by:
\begin{rosi}
histo : forall z f. (Xh z f) -> Mu (Sigma z) -> f z
histo = \ a. fix (\ rec x. a (out x) rec)
\end{rosi}
An !Xh! algebra !Xh z f! determines a transformation from inputs of type !Mu (Sigma z)! to result of type !f z!. There are three significant components of extensible histomorphisms.

First, rather than either performing the traversal bottom-up, as catamorphisms do, or providing unfettered access to recursive subdata, as a normal recursive definition does, recursive subdata is given a generic type (here !Mu (Sigma w)! for a fresh type variable !w!), and can only be manipulated through a provided ``abstract recursive call'' function !Mu (Sigma w) -> f w!. This allows the !Xh! algebra to decide when and whether it will make recursive calls, but still abstracts over the structure of those calls. Mendler used this idea to guarantee termination. For us, it guarantees that recursive calls will remain well-typed as we combine algebras.

Second, we provide some access to the structure of recursive subdata. Rather than leaving !w! unconstrained, as would be typical for Mendler-style recursion schemes, we provide that !z! is contained in !w!. This even allows deep pattern matching, as would be needed for small-step evaluation, so long as a default for cases outside !z! can be provided. (The typical approach for Mendler-style histomorphisms would be to keep the recursive type fully abstract, say !r!, but provide an ``abstract \Tt{out}'' function !r -> Sigma w r!, but this approach is too limiting to define closures.)

Third, we allow the output type to vary with the input type. This allows us to describe closures (the values produced from the functional language), where the captured expression may use any of the expressions in the input type. %

Finally, we need to describe how !Xh! algebras can be combined. Here is the fully explicit account:
\begin{rosi}
orXh : forall x y z f. x + y ~ z => Xh x f -> Xh y f -> Xh z f
orXh = \ a b. /\ w. a [w] | b [w]
\end{rosi}
!orXh a b! is itself an !Xh! algebra, so it is defined over an abstract row !w!. We bring that row into scope, and then provide it as the abstract row required by !a! and !b!. For !a!, we must then have that !x < w!, but !x < z! follows from the assumption !x + y ~ z!, and then !x < w! follows from transitivity. A parallel argument holds for !b!. In practice, our implementation successfully infers the type abstractions and applications, so we can combine !Xh! algebras directly with !(|)!.

\subsection{Example: Big-step Evaluation}
\label{sec:big-step}

As an example of using extensible histomorphisms, we develop big-step evaluation functions for each sublanguage, and combine them to get a big-step evaluation function for the entire language. Our big-step functions will take terms to mappings from environments to values. We need several auxiliary definitions. The !Env! type function maps the structure of (recursive) values to environments of those values:
\begin{rosi}
type Env : R[* -> *] -> *
type Env = \z. List (Mu (Sigma z))
\end{rosi}
The !as! operator will be used to pipeline failure in evaluation, and has something of the flavor of the monadic bind operator:
\begin{rosi}
as : forall l f z. { l := f } < z, { 'Err := (\ w. Unit) } < z =>
     #l -> Mu (Sigma z) -> (f (Mu (Sigma z)) -> Mu (Sigma z)) -> Mu (Sigma z)
as = \ l w f. match (out w) ( case l f | const (rcon #'Err tt) )
\end{rosi}

Now we can define the big-step evaluation function for the arithmetic language. We begin with a row !ValA! which gives the shape of arithmetic values.
\begin{rosi}
type ValA : R[R[* -> *] -> * -> *]
type ValA = { 'Nat := (\ expr. \ val. Nat), 'Err := (\ expr. \ val. Unit) }
\end{rosi}
Each case in !ValA! takes two parameters. The first is a row giving the shape of expressions. The second is the recursive parameter. Neither are necessary for arithmetic. The only values we require are natural numbers and errors. Evaluation itself is given by  an !Xh! algebra over the !ArithF! row:
\begin{rosi}
evalA : forall valr. ValA < valr =>
        Xh ArithF (\ expr. Env (valr expr) -> Mu (Sigma (valr expr)))
evalA = \ exp rec env.
           match exp
           ( case #'IConst (rcon #'Nat)
           | case #'Plus (\ p.
               as #'Nat (rec (sel p #'1) env) (\ i.
               as #'Nat (rec (sel p #'2) env) (\ j.
                 rcon #'Nat (add i j)))))
\end{rosi}
The result type of the histomorphism is a function from environments to values. The structure of values itself is also extensible: !evalA! is defined for any value row !valr! that contains the !ValA! cases needed for arithmetic. The result of the histomorphism is defined in terms of the full input type !expr!, which contains !ArithF! but may have many more cases. We use this final expression type in constructing values. That is, !valr expr! may refer to both a value type larger than !ValA! and an expression type larger than !ArithF!.

The interesting case of !evalA! is for addition. Intuitively, we want to recursively evaluate the subexpressions of the addition, and then sum the results. The complication is that we cannot be sure that the recursive evaluation calls will return natural number values---as !evalA! is polymorphic in the description of values, the results of recursive calls may have more constructors than just !'Nat! and !'Err!. We handle these potential failures using the !as! auxiliary function.

The big-step function for the Boolean language !evalB! is structured identically to that for arithmetic, and is given in \Appendix{sec:extra-code-samples}.

Next we give the big-step function for the functional language. Values are closures, represented as a pair of an environment (that is, a list of values) and an expression:
\begin{rosi}
type ValL : R[R[* -> *] -> * -> *]
type ValL = { 'Clos := (\ expr. \ val. Pair (List val) (Mu (Sigma expr)))
            , 'Err := (\ expr. \ val. Unit) }
\end{rosi}

Here, at last, we need all the structure we have built up for values. The first component is defined in terms of the recursive parameter !val!. The second component is defined in terms of the structure of expressions !expr!. As in the previous cases, the big-step function itself is a histomorphism from !LamF! expressions to environment to value mappings, where the structure of values must contain !ValL!:
\begin{rosi}
evalL : forall valr. ValL < valr => Xh LamF (\ expr. Env (valr expr) -> Mu (Sigma (valr expr)))
\end{rosi}
In a closure value, then, the environment may contain values other than those described by !ValF!, and the expression itself may contain constructs other than those described by !LamF!.

The cases of the !evalL! function itself should be unsurprising:
\begin{rosi}
evalL = \ exp rec env.
           match exp
           ( case #'Var (\ x. fromMaybe (rcon #'Err tt) (nth env x))
           | case #'Lam (\ e. rcon #'Clos (pair env e))
           | case #'App (\ p.
               as #'Clos (rec (sel p #'1) env) (\ clos.
                 (rec (sel clos #'2)) (cons (rec (sel p #'2) env) (sel clos #'1)))))
\end{rosi}
The !nth!, !fromMaybe!, and !pair! functions in the first two cases behave as their names would suggest. In the third case, we evaluate the function to get a closure, and evaluate the body of the closure in a suitably extended environment.

Finally, we tie the previous functions together to get a big-step function for the entire language. Due to limitations in our implementation, we must explicitly define the combined language !AllF!.
\begin{rosi}
type AllF : R[* -> *]
type AllF = { 'BConst := (\ t. Bool), 'If := (\ t. Triple t t t)
            , 'IConst := (\ t. Nat), 'Plus := (\ t. Pair t t)
            , 'Var := Const Nat, 'Lam := Id, 'App := (\ t. Pair t t) }
\end{rosi}
The evaluation function for the combined language is an !Xh! algebra, just as the evaluation functions for the individual languages. It requires a value structure that contains the requirements of each of its cases, however we can still leave the door open to more extension. The evaluation function is defined simply by branching among the smaller evaluation functions:
\begin{rosi}
eval : forall valr. ValA < valr, ValB < valr, ValL < valr =>
       Xh AllF (\ expr. Env (valr expr) -> Mu (Sigma (valr expr)))
eval = evalA | evalB | evalL
\end{rosi}

This example demonstrates freely combinable types and freely combinable functions, with results that vary with their inputs. The only significant challenge in writing this example was correctly describing the structure of value rows, and the polymorphisms over value rows. Otherwise, implementing the individual functions was appealingly direct.

\subsection{Example: Desugaring}
\label{sec:desugaring}

For a final example of extensible recursive programming, we consider desugaring one of our sublanguages into another. Our goal is to desugar the Boolean language into Church-encodings of Booleans in the functional language. (Of course, we are at this point relying on our language being untyped.) The challenge is that this desugaring should be defined only in terms of the Boolean and functional langauges. It should not matter what other language features are included, but Boolean expressions must be desugared uniformly throughout.

We desugar Boolean expressions by:
\begin{rosi}
desugarB : forall z. LamF < z => Xh BoolF (\ w. Mu (Sigma z))
desugarB = \ exp rec.
            match exp
            ( case #'BConst
               ( case #'True (\ u. rcon #'Lam (rcon #'Lam (rcon #'Var one)))
               | case #'False (\ u. rcon #'Lam (rcon #'Lam (rcon #'Var zero))))
            | case #'If (\ t. app (app (rec (fst t)) (rec (snd t))) (rec (thd t))))
\end{rosi}
Desugaring is defined as an !Xh! algebra from !BoolF! to any expression that that includes !LamF!. As expected, we interpret !True! as $\lambda x. \lambda y. x$, !False! as $\lambda x. \lambda y. y$, and conditionals as application. The !app! convenience function handles the mechanics of building application expressions:
\begin{rosi}
app : forall z. LamF < z => Mu (Sigma z) -> Mu (Sigma z) -> Mu (Sigma z)
app = \ f e. rcon #'App (pair f e)
\end{rosi}
!one! and !zero! are the expected numerals, and in transforming conditionals we begin by recursively desugaring the subexpressions.

We extend desugaring from Boolean to arbitrary expressions:
\begin{rosi}
ext : forall z : R[* -> *]. Pi (Functor z) -> Xh z (\w. Mu (Sigma z))
ext = \ d v rec. in (fmapS d rec v)
\end{rosi}
\codesep
\begin{rosi}
desugar : forall y. BoolF < y, LamF < y - BoolF =>
          Pi (Functor (y - BoolF)) -> Xh y (\w. Mu (Sigma (y - BoolF)))
desugar = \ d. desugarB | ext d
\end{rosi}
The type of !desugar! captures its intended operation: it is an algebra from expressions over !y! to expressions over !y - BoolF!. We require that !y! contain Boolean expressions, for !y - BoolF! to be meaningful, and that !y - BoolF! contain functional expressions, for the desugaring to be meaningful. The interesting question is how we will handle the non-!BoolF! cases of !y!, given that !y! is otherwise arbitrary. Our solution is given away in the type: we require a record of !fmap! functions for the !y - BoolF! cases, with which we can make the required recursive transformations. We extract the generic logic to !ext!, an algebra that does nothing. Essentially, however, it traverses its argument while doing nothing at each case. Finally, we describe the generic desugaring function by combining the logic for Booleans !desugarB! with the generic traversal !ext!.

This example shows a different application of extensibility: modularity. The !desugar! function could be viewed to be complete (although, as written, it could still be further extended.) Nevertheless, extensibility has still played two roles. First, it has allowed us to write the function without having to fix the remaining cases, or list them in our implementation. More importantly, because the function is defined over arbitrary expressions, a user of this function can be sure that it will not manipulate any cases other than the !BoolF! ones, and, because !BoolF! does not appear in the result, that those cases will be completely desugared.

\subsection{Not All Functions are Extensible Histomorphisms}

We find programming with algebras to be a necessity. Troubles arise when trying to combine histomorphisms (of type !Mu (Sigma z) -> f z!) rather than their constituent !Xh! algebras. Consider the combination of two variant eliminators !f : Mu (Sigma x) -> a! and !g : Mu (Sigma y) -> a! into a third at type !Mu (Sigma z) -> a!, provided !x + y ~ z!. Suppose we even know that the variant we have at type !Mu (Sigma z)! was built from a label residing in !x!. After using !out! to receive data at type !z (Mu Sigma z)!, we can demote it to type !x (Mu Sigma z)!. But here we are stuck: its subdata is still at the "larger" type !Mu (Sigma z)!, which we cannot possibly demote to type !Mu (Sigma x)!. Hence !f! appears useless to us.

Alternatively, we might look to "promote" !f! to handle (by means of a default traversal case) the larger type !Mu (Sigma z)!. That is, we might hope to extract an extensible histomorphic algebra from the shape of !f!, essentially by identifying and abstracting over any recursive calls in its definition. Of course, some subtleties will remain. For example, the programmer writing the !evalA! and !evalB! functions would still have to make the result type sufficiently polymorphic, but this is no different than the problem the programmer faces writing the !Xh! algebras. The relationship between the input and output type of !evalF! might be difficult to discover automatically, but again one could hope for clever techniques to work in most cases.

The deeper problem with this approach is simply that not all functions over recursive variants need to be extensible histomorphisms. Consider the !ext! transformation \cref{sec:desugaring}. If we erase the distinction between extensible histomorphisms and arbitrary functions, the type of !ext! is !forall z : R [* -> *]. Pi (Functor z) -> Mu (Sigma z) -> Mu (Sigma z)!. This type is equally well inhabited by !const id!. However, !const id! cannot play the role of !ext! in defining !desugar!, because it does not describe a traversal at all.

Of course, this is not to say that syntactic sugar could not make writing extensible histomorphisms more pleasant, or at least initially more familiar. Whatever their presentation, however, we claim that some distinction between extensible histomorpisms and arbitrary functions must remain.

\InlineOff
\section{Metatheory}
\label{sec:metatheory}

This section outlines the metatheory of \Rome, via small-step operational semantics for the type, evidence, and term languages. Our metatheory is mechanized in Agda, and the claims of this section are annotated with the corresponding points in our mechanized development.

\subsection{Type reduction}
\label{sec:type-reduction}

\begin{figure}
\begin{gather*}
\begin{array}{r@{\hspace{7px}}l@{\qquad\qquad}r@{\hspace{7px}}l}
  \text{Type variables} & \alpha \in \mathcal A &
  \text{Labels} & \ell \in \mathcal L
\end{array} \\
\begin{doublesyntaxarray}
  \mcl{\text{Ground Kinds}}  & \gamma   & ::= & \TypeK \mid \LabK \\
  \mcl{\text{Kinds}}         & \kappa    & ::= & \gamma \mid \kappa \to \kappa \mid  \RowK \kappa \\
\end{doublesyntaxarray}
\qquad
\begin{doublesyntaxarray}
  \mcl{\text{Row Literals}}   & \NormalRows \ni \Normal \rho    & ::= & \RowIx i 0 m {\LabTy {\ell_i} {\Normal {\tau_i}}} \\
  \mcl{\text{Neutral Types}} & n    & ::= & \alpha \mid n \, {\Normal \tau}  \\
\end{doublesyntaxarray}
\\
\begin{doublesyntaxarray}
  \mcl{\text{Normal Types}}  & \NormalTypes \ni \Normal \tau, \Normal \upsilon, \Normal \phi & ::= & n \mid \hat{\phi}^{\star} \, n \mid \Normal{\rho} \mid \Normal{\pi} \then \Normal{\tau} \mid \forall \alpha\co\kappa. \Normal{\tau} \mid \lambda \alpha\co\kappa. \Normal{\tau} \\
                             &       &     & \mid & \LabTy n {\Normal \tau} \mid \ell \mid \Sing {\Normal \tau} \mid {\Normal \tau} \Compl {\Normal \tau} \mid \KFam \Pi \TypeK \, {\Normal \tau} \mid \KFam \Sigma \TypeK \, {\Normal \tau}  \\
\end{doublesyntaxarray}
\end{gather*}
\begin{small}
\begin{gather*}
\fbox{$\KindJNF \Delta {\Normal \tau} {\kappa}$} \, \fbox{$\KindJNE \Delta {n} {\kappa}$} \\
\ib{
  \irule[\kruleNF{ne}]
    {\KindJNE \Delta n \gamma};
    {\KindJNF \Delta n \gamma}}
\isp
\ib{
  \irule[\kruleNF{$\Compl$}]
    {\KindJNF \Delta {\Normal {\tau_i}} {\RowK \kappa}}
    {\Normal{\tau_1} \notin \NormalRows \, \text{or}\, \Normal{\tau_2} \notin \NormalRows};
    {\KindJNF \Delta {\Normal{\tau_2} \Compl \Normal{\tau_1}} {\RowK \kappa}}}
\isp
\ib{
  \irule[\kruleNF{$\triangleright$}]
    {\KindJNE \Delta {n} {\LabK}}{\KindJNF \Delta {\Normal \tau} \kappa};
    {\KindJNF \Delta {\LabTy n {\Normal \tau}}{\RowK \kappa}}}
\end{gather*}
\end{small}
\caption{Normal type forms}
\label{fig:type-normalization}
\end{figure}

We define reduction on types $\tau \RedT \tau'$ by directing the type equivalence judgment $\TEqvJ \varepsilon \tau {\tau'} \kappa$ from left to right (with the exception of rule \errule{map$_\mathsf{id}$}, which reduces right-to-left). Importantly, every type has a normal form, which we exhibit in our mechanization via evaluation, largely following the techniques employed by \citeposs{ChapmanKNW19} \emph{normalization by evaluation} of types in System F$\omega\mu$.

The syntax of neutral and normal types is given in \figref{type-normalization}. We use hats to denote normalized sets of syntax, e.g., we write $\Types$ for the set of well-kinded types and $\NormalTypes$ for the set of well-kinded normal types. (As our mechanized semantics is intrinsically typed, it is convenient to consider only sets of well-kinded syntax.) When possible and convenient, we use hats on meta-variables (e.g., $\Normal \upsilon$) to obviate that we are referring to normal entities. The syntax of predicates $\Normal \pi$ containing normal types is as expected and omitted.

Normalization reduces applications and maps except when a variable blocks computation, which we represent as a \emph{neutral type}. A neutral type is either a variable or a spine of applications with a variable in head position. We distinguish ground kinds $\gamma$ from functional and row kinds, as neutral types may only be promoted to normal type at ground kind (rule \kruleNF{ne}): neutral types $n$ at functional kind must $\eta$-expand to have an outer-most $\lambda$-binding (e.g., to $\lambda x. \, n\, x$), and neutral types at row kind are expanded to an inert map by the identity function (e.g., to $\Map {(\lambda x. x)} \, n$). Likewise, repeated maps are necessarily composed according to rule \errule{map$_\circ$}: For example, $\Map \phi_{1}\, (\Map \phi_{2}\, n)$ normalizes by letting $\phi_{1}$ and $\phi_{2}$ compose into $(\Map {(\phi_{1} \circ \phi_{2})}\, n$). By consequence of $\eta$-expansion, records and variants need only be formed at kind $\TypeK$. This means a type such as $\Pi (\LabTy {\ell} {\lambda x. x})$ must reduce to $\lambda x. \Pi (\LabTy {\ell} {x})$, $\eta$-expanding its binder over the $\Pi$. Nested applications of $\Pi$ and $\Sigma$ are also "pushed in" by rule \erule{$\Xi$}. For example, the type $\Pi \, \Sigma \, (\LabTy {\ell_1} {(\LabTy  {\ell_2} \tau)})$ has $\Sigma$ mapped over the outer row, reducing to $\Pi (\LabTy {\ell_1} {\Sigma (\LabTy {\ell_2} \tau)})$.

The syntax $\LabTy n \Normal \tau$ separates singleton rows with variable labels from row literals $\Normal \rho$ with literal labels; rule \kruleNF{$\triangleright$} ensures that $n$ is a well-kinded neutral label.  A row is otherwise an inert map $\Map \phi \, n$ or the complement of two rows $\Normal{\tau_2} \Compl \Normal{\tau_1}$. Observe that the complement of two row literals should compute according to rule \erule{$\Compl$}; we thus require in the kinding of normal row complements \kruleNF{$\Compl$} that one (or both) rows are not literal so that the computation is indeed inert. The remaining normal type syntax does not differ meaningfully from the type syntax; the remaining kinding rules for the judgments $\KindJNF \Delta {\Normal \tau} \kappa$ and $\KindJNE \Delta n \kappa$ do not differ meaningfully from their analogues in \figref{kinding}.

We carefully define the normal type syntax so that no type $\Normal \tau \in \NormalTypes$ could reasonably reduce further to some other $\tau' \in \NormalTypes$. Hence we write $\tau \NRedT$ synonymously with $\tau \in \NormalTypes$ to indicate that $\tau$ is well-kinded and has no further reductions. We define a normalization function in Agda to materialize this sentiment.

\begin{theorem}[Normalization] ~
  There exists a normalization function $\Norm\, : \Types \to \NormalTypes$\footnote{\CiteAgda{Rome.Types.Semantic.NBE}{$\Norm$}} that maps well-kinded types to well-kinded normal forms.
\end{theorem}

$\Norm$ is realized in Agda intrinsically as a function from derivations of $\KindJ \Delta \tau \kappa$ to derivations of $\KindJNF \Delta {\Normal{\tau}} \kappa$.  Conversely, we witness the inclusion $\NormalTypes \subseteq \Types$ as an embedding $\Embed \,:\, \NormalTypes \to \Types$\footnote{\CiteAgda{Rome.Types.Normal.Syntax}{$\Embed$}}, which casts derivations of $\KindJNF \Delta {\Normal \tau} \kappa$  back to a derivation of $\KindJ \Delta \tau \kappa$; we omit this function and its use in the following claims, as it is effectively the identity function (modulo tags).

The following properties confirm that $\Norm$ behaves as a normalization function ought to. The first property, \emph{stability}, asserts that normal forms cannot be further normalized. Stability implies \emph{idempotency} and \emph{surjectivity}.

\begin{theorem}[Properties of normalization] ~
  \begin{itemize}
  \item (Stability) for all $\Normal\tau \in \NormalTypes$, $\Norm \, \Normal\tau = \Normal\tau$.\footnote{\CiteAgda{Rome.Types.Theorems.Stability}{stability}}
  \item (Idempotency) For all $\tau \in \Types$, $ \Norm (\Norm \, \tau) =\, \Norm \, \tau$.\footnote{\CiteAgda{Rome.Types.Theorems.Stability}{idempotency}}
  \item (Surjectivity) For all $\Normal\tau \in \NormalTypes$, there exists $\upsilon \in \Types$ such that $\Normal\tau =\, \Norm \upsilon$.\footnote{\CiteAgda{Rome.Types.Theorems.Stability}{surjectivity}}
  \end{itemize}
\end{theorem}

We now show that $\Norm$ indeed reduces faithfully according to the equivalence relation $\TEqvJ \Delta \tau \tau \kappa$. Completeness of normalization states that equivalent types normalize to the same form.

\begin{theorem}[Completeness]
For well-kinded $\tau , \upsilon \in \Types$ at kind $\kappa$, If $\TEqvJ \Delta \tau \upsilon \kappa$ then $\Norm\, \tau = \, \Norm \, \upsilon$.\footnote{\CiteAgda{Rome.Types.Theorems.Completeness}{completeness}}
\end{theorem}

\Ni Soundness of normalization states that every type is equivalent to its normalization.

\begin{theorem}[Soundness]
For well-kinded $\tau \in \Types$ at kind $\kappa$, there exists a derivation that $\TEqvJ \Delta \tau {\,\Norm \tau} {\kappa}$.\footnote{\CiteAgda{Rome.Types.Theorems.Soundness}{soundness}} Equivalently\footnote{\CiteAgda{Rome.Types.Theorems.Soundness}{Soundness$\to$Completeness$^{-1}$,Completeness$^{-1}$$\to$Soundness}}, if $\Norm\, \tau =\, \Norm\, \upsilon$, then $\TEqvJ \Delta \tau \upsilon {\kappa}$.\footnote{\CiteAgda{Rome.Types.Theorems.Soundness}{completeness$^{-1}$}}
\end{theorem}
\Ni Soundness and completeness together imply, as desired, that $\tau \RedT \tau'$ iff $\Norm\, \tau =\, \Norm \tau'$.

Equivalence of normal types is syntactically decidable which, in conjunction with soundness and completeness, is sufficient to show that \Rome's equivalence relation is decidable. Consequently, the user has no obligation to provide proofs of equivalence in type and predicate conversion (rules \trule{conv} and \entrule{conv}).

\begin{theorem}[Decidability]
  Given well-kinded $\tau, \upsilon \in \Types$ at kind $\kappa$, the judgment $\TEqvJ \Delta \tau \upsilon \kappa$ either (i) has a derivation or (ii) has no derivation.\footnote{\CiteAgda{Rome.Types.Normal.Properties.Decidability}{\_$\equiv$t?\_}}
\end{theorem}

The normal type syntax is pleasantly partitioned by kind. Due to $\eta$-expansion of functional variables, arrow kinded types are canonically $\lambda$-bound. A normal type at kind $\RowK \kappa$ is either an inert map $\hat{\phi}^{\TypeK} \, n$, a variable-labeled row $(\LabTy n \Normal \tau)$, the complement of two rows $\Normal {\tau_{2}} \Compl \Normal {\tau_{1}}$, or a row literal $\Normal \rho$. The first three cases necessarily have neutral types (recall that at least one of the two rows in a complement is not a row literal). Hence rows in empty contexts are canonically literal. Likewise, the only types with label kind in empty contexts are label literals; recall that we disallowed the formation of $\Pi$ and $\Sigma$ at kind $\RowK \LabK \to \LabK$, thereby disallowing non-literal labels such as $\Delta \not\vdash \Pi \epsilon \co \LabK$ or $\Delta \not\vdash \Pi (\LabTy {\ell_{1}} {\ell_{2}}) \co \LabK$.

\begin{theorem}[Canonicity]
  Let $\Normal\tau \NRedT$.
  \begin{itemize}
    \item If $\KindJNF \Delta {\Normal\tau} {(\kappa_{1} \to \kappa_{2})}$ then $\Normal\tau = {\lambda \alpha \co \kappa_{1}. \Normal\upsilon}$;\footnote{\CiteAgda{Rome.Types.Normal.Syntax}{arrow-canonicity}}
    \item if $\KindJNF \epsilon {\Normal\tau} {\RowK \kappa}$ then $\Normal\tau =  \RowIx i 0 m {\LabTy {\ell_i} {\Normal{\tau_i}}}$.\footnote{\CiteAgda{Rome.Types.Normal.Syntax}{row-canonicity-$\emptyset$}}
    \item If $\KindJ \epsilon {\Normal\tau} \LabK$, then $\Normal\tau = \ell$.\footnote{\CiteAgda{Rome.Types.Normal.Syntax}{label-canonicity-$\emptyset$}}
  \end{itemize}
\end{theorem}

\subsection{Runtime Syntax}

\begin{figure}
\begin{syntax}
  \text{Runtime expressions} & R & ::= & \dots \mid \Record [\rho] {\vec R} \mid \Variant [\rho] i R \\
  \text{Values} &\NormalTerms \ni V & ::= & \Lambda \alpha \co \kappa. M \mid \Lambda v \co \pi. M \mid \lambda x \co {\Normal \tau}. M
  \mid \In \, V \\
  & & \mid & \Sing {\Normal \tau} \mid \Record [\RowIx i 0 m {\LabTy {\xi_i} {\tau_i}}] {\vec V} \mid \Variant [\RowIx j 0 m {\LabTy {\xi_j} {\tau_j}}] i V \\
  & \NormalEvidence \ni V_Q & ::= & \InclV \, p \mid \CombV \, p \, q \\
  \text{Constant spines} & S & ::= & K \mid \AppT S \tau \mid S \AppE {V_Q} \mid S \, V \\
  \text{Evaluation contexts} & E & ::= & \square \mid E \, N \mid \AppT E \tau \mid \AppT E Q \mid S \, E \\
  & & \mid & \LabTermX \Xi E N \mid \LabTermX \Xi M E \mid \UnlabelX \Xi E N \mid \UnlabelX \Xi M E \qquad \qquad \Xi \in \Set {\Pi, \Sigma} \\
  & & \mid & \Record [\rho] {V_1, \dots, V_n, E, R_1, \dots, R_m} \mid \Variant [\rho] i E \\
  & E_{\mathcal Q} & ::= & \TransV {E_{\mathcal Q}} {Q} \mid \TransV {V_Q} {E_{\mathcal Q}} \mid \LeqV {map} \, E_{\mathcal Q} \mid \LeqV {plusL} \, E_{\mathcal Q} \mid \LeqV {plusR} \, E_{\mathcal Q} \\
  & & \mid & \PlusV {map} \, E_{\mathcal Q} \mid \PlusV {complL} \, E_{\mathcal Q} \mid \PlusV {complR} \, E_{\mathcal Q}
\end{syntax}
\begin{small}
\begin{gather*}
\fbox{$\TypeJ \Delta \Phi \Gamma R \tau$}
\\
\ib{\irule[\trunrule{$\Pi$}]
          {\TypeJ \Delta \Phi \Gamma {R_i} {\tau_i}};
          {\TypeJ \Delta \Phi \Gamma
                  {\Record [\RowIx i 0 m {\LabTy {\xi_i} {\tau_i}}] {{\vec R}}}
                  {\Pi {\RowIx i 0 m {\LabTy {\xi_i} {\tau_i}}}}}}
\isp
\ib{\irule[\trunrule{$\Sigma$}]
          {\TypeJ \Delta \Phi \Gamma R {\tau_i}};
          {\TypeJ \Delta \Phi \Gamma
                  {\Variant [\RowIx j 0 m {\LabTy {\xi_j} {\tau_j}}] i R}
                  {\Sigma \RowIx j 0 m {\LabTy {\xi_j} {\tau_j}}}}}
\end{gather*}
\end{small}
\caption{Runtime syntax}
\label{fig:syntax-runtime}
\end{figure}

\cref{fig:syntax-runtime} describes additional syntax to capture evaluation. We extend the grammar of expressions with literals for records $\Record [\rho] {\vec R}$ and variants $\Variant [\rho] i R$. We write $\vec R$ for a vector of terms, and write $\vec R_i$ for its $i^{\text{th}}$ entry. These terms have the expected typing rules. Values are abstractions (over types, evidence, and terms), singleton constants, and record and variant literals. We assume that constants will only appear fully applied, which can be assured with trivial $\eta$-expansion. Evidence values are $\InclV \, p$ (for containment predicates) and $\CombV \, p \, q$ (for combination predicates). We define evaluation contexts for both terms ($E$) and evidence ($E_{\mathcal Q}$). A spine $S$ is a constant applied to a sequence of types, evidence values, or term values, and is used to define contexts with constant applications.

\subsection{Evidence Functions}

To define the evaluation of the constants, we need two helper functions on evidence values. We describe these function here, and give their implementations in \Appendix{sec:extra-evaluation}. The first is used in interpreting combination evidence $\CombV \, p \ q : \PlusP x y z$. In this evidence, $p$ is evidence of $x$'s containment in $z$, and $q$ is evidence of $y$'s containment in $z$. However, to implement record concatenation and variant branching, we need a mapping from entries in $z$ to entries in either $x$ or $y$, which we call $\mathsf{pick} : (\mathbb N \pto \mathbb N) \times (\mathbb N \pto \mathbb N) \times \mathbb N \pto \mathbb N + \mathbb N$. In practice, and indeed in our mechanized semantics, the $\mathsf{pick}$ map is computed in entailment and included in combination evidence. The second is a function that computes the ``dual'' of an evidence value $\InclV \, p : \LeqP x z$. That is to say: $\mathsf{dual} : (\mathbb N \pto \mathbb N) \to \mathbb N \pto \mathbb N$ computes from $p$ an inclusion map for all the entries of $z$ not in the range of $p$; because inclusion maps must be increasing, this inclusion map is unique.

\subsection{Operational Semantics: Functional Terms}

\cref{fig:red} gives reduction rules for the functional and recursive terms of \Rome, which are standard. We depend on substitution for types\footnote{\CiteAgda{Rome.Types.Substitution}{sub$_{k}$}}\footnote{\CiteAgda{Rome.Types.Normal.Substitution}{sub$_{k}$NF}}, terms\footnote{\CiteAgda{Terms.Normal.Substitution}{sub}}, and evidence\footnote{\CiteAgda{Terms.Normal.Substitution}{subEnt}}, which are all defined in an entirely standard way. \Ruleref{cong-type} reduces types when they appear as arguments to constants.

\begin{figure}
\begin{small}
\begin{minipage}[t]{0.44\linewidth}
\begin{align}
  \AppT {(\Lambda \alpha\co\kappa. M)} \tau &\Red M[\tau/\alpha]
  \label{rule:beta-forall}
  \tag{\rbrule\forall}
  \\
  \AppT {(\Lambda v\co\pi. M)} Q &\Red M[Q/v]
  \label{rule:beta-then}
  \tag{\rbrule\then}
  \\
  (\lambda x\co\tau. M) \, N &\Red M[N/x]
  \label{rule:beta}
  \tag{\rbrule\to}
  \\
  \Out \, {(\In \, V)} &\Red V
  \label{rule:out-in}
  \tag{\rdrule\mu}
\end{align}
\end{minipage}
\hspace{0.1\linewidth}
\begin{minipage}[t]{0.44\linewidth}
\begin{align}
  \Fix \, f &\Red f \, (\Fix \, f)
  \label{rule:fix}
  \tag{\rdrule{\mathsf{fix}}}
  \\
  E[M] &\Red E[M'] \qquad \qquad \text{if $M \Red M'$}
  \label{rule:cong}
  \tag{$\xi$}
  \\
  E[S \, \tau] &\Red E[S \, \tau']
    \qquad \qquad \text{if $\tau \RedT \tau'$}
  \label{rule:cong-type}
  \tag{$\xi_{\mathcal T}$}
\end{align}
\end{minipage}
\end{small}
\caption{Reduction: Functions and Recursion}
\label{fig:red}
\end{figure}

\subsection{Operational Semantics: Variants}

\begin{figure}
\begin{small}
\begin{align}
  \LabTermX \Sigma {\Sing \ell} V &\Red \Variant [\Row {\LabTy \ell \tau}] 0 V
  \label{rule:label-variant}
  \tag{\rdrule{\triangleright\Sigma}}
  \\
  \UnlabelX \Sigma {\Variant [\Row {\LabTy \ell \tau}] 0 V} {\Sing \ell} &\Red V
  \label{rule:unlabel-variant}
  \tag{\rdrule{/\Sigma}}
  \\
    \AppT {\Inj} {\Normal\rho_y,\Normal\rho_z,\InclV \, p} \, \Variant [\rho_y] i V &\Red \Variant [\rho_z] {p(i)} V
  \label{rule:inj}
  \tag{\rdrule{\mathsf{inj}}}
  \\
  \AppT {(\Branch)} {\Normal\rho_x,\Normal\rho_y,\Normal\rho_z,\CombV \, p \, q} \, f \, g \, \Variant [\rho_z] i V &\Red
    \begin{cases}
      f \, \Variant [\rho_x] j V &\text{if $\mathsf{pick}(p, q, i) = \mathsf{Left} \, j$} \\
      g \, \Variant [\rho_y] j V &\text{if $\mathsf{pick}(p, q, i) = \mathsf{Right} \, j$}
    \end{cases}
  \label{rule:branch}
  \tag{\rdrule{{\Branch}}}
  \\
  \AppT {\Ana} {\phi, \RowIx j 0 m {\LabTy {\ell_j} \Normal\tau_j}, {\Normal\upsilon}} \, \Sing \phi \, V \, \Variant [\RowIx j 0 m {\LabTy {\ell_j} {\tau_j'}}] i W &\Red
    \AppT V {\ell_i,\Normal\tau_i,\InclV \, (0 \mapsto i)} \, \Sing {\ell_i} \, W
  \label{rule:ana}
  \tag{\rdrule{\mathsf{ana}}}
\end{align}
\end{small}
\caption{Reduction: Variants}
\label{fig:red-var}
\end{figure}

\Cref{fig:red-var} gives reduction rules for variants. \Ruleref{label-variant} constructs singleton variants---because we know the constructed variant will be a singleton, we can be sure the constructor label is at position 0 in its type. \Ruleref{unlabel-variant} eliminates singleton variants. The alignment of the label $\ell$ in the type annotation on the variant value and the singleton $\Sing \ell$ argument is guaranteed for well-typed terms.

\Ruleref{inj} injects smaller variants into larger variants; the evidence $p$ determines the position of the constructor $i$ in the larger variant. \Ruleref{branch} implements variant branching. If constructor $i$ is entry $j$ in $\rho_x$, the $f$ case is invoked with the new constructor, and similarly for the $g$ case. \Ruleref{ana} implements analysis. We must invent evidence that $\Row {\LabTy {\ell_i} {\tau_i}}$ is contained in $\RowIx j 0 m {\LabTy {\ell_j} {\tau_j}}$. But this must always be exactly the index used to construct the final argument to \lstinline!ana!.

In all of these steps, type information is used solely to construct other type annotations, never to determine the meaning of terms. The one case where this may seem untrue is the singleton $\Sing {\ell_i}$ on the right-hand side of \ruleref{ana}, but as $\Sing \ell$ is a singleton type, this is essentially a unit value.

\subsection{Operational Semantics: Records}

\begin{figure}
\begin{small}
\begin{align}
  \LabTermX \Pi {\Sing \ell} V &\Red \Record [\Row {\LabTy \ell \tau}] {V}
  \label{rule:label-record}
  \tag{\rdrule{\triangleright\Pi}}
  \\
  \UnlabelX \Pi {\Record [\Row {\LabTy \ell \tau}] V} {\Sing \ell} &\Red V
  \label{rule:unlabel-record}
  \tag{\rdrule{/\Pi}}
  \\
  \AppT {\Prj} {\Normal{\rho_y}, \Normal{\rho_z}, \InclV \, p} \, \Record [\rho_z] {\vec V} &\Red
    \Record [\rho_y] {V_{p(i)} \mid i \in 0 \dots |\Normal{\rho_y}|}
  \label{rule:prj}
  \tag{\rdrule{\Prj}}
  \\
  \AppT {(\Concat)} {\Normal{\rho}_x,\Normal{\rho}_y,\Normal{\rho}_z,\CombV\,p \, q} \, \Record [\rho_x] {\vec V} \, \Record [\rho_y] {\vec W} &\Red
    \left\langle
      \left.
      \begin{cases}
        V_j &\text{if $\mathsf{pick}(p, q, i) = \mathsf{Left} \, j$} \\
        W_j &\text{if $\mathsf{pick}(p, q, i) = \mathsf{Right} \, j$}
      \end{cases}
      \right\vert
      i \in 0 \dots |\Normal\rho_z|
    \right\rangle%
  \label{rule:concat}
  \tag{\rdrule{\Concat}}
  \\
  \AppT {\Syn} {\phi,\RowIx j 0 m {\LabTy {\ell_j} \Normal\tau_j}} \, {\Sing \phi} \, V &\Red
    \Record [\RowIx j 0 m {\LabTy {\xi_i} {\upsilon_i}}] {\AppT V {\ell_i, \Normal\tau_i, \InclV \, (0 \mapsto i)} \, \Sing {\ell_i} \mid i \in 0..m} %
  \label{rule:syn}
  \tag{\rdrule{\mathsf{syn}}}
\end{align}
\end{small}
\caption{Reduction: Record terms}
\label{fig:red-rec}
\end{figure}

\Cref{fig:red-rec} gives reduction rules for records. Rules (\ref{rule:label-record}) and (\ref{rule:unlabel-record}) construct and deconstruct singleton records. In the following rules, we use the syntax $\Record [\rho] {V_i \mid i \in 0 \dots m}$ to denote the record $\Record [\rho] {V_0, V_1, \dots, V_m}$, and $|\rho|$ to denote the size (that is, the number of entries) in $\rho$. As we reduce types as arguments to constants, we can always be sure that we have row literals and so that this operator is well-defined.

\Ruleref{prj} projects smaller records from larger records; the evidence $p$ finds each field of the smaller record in the larger one. \Ruleref{concat} concatenates records, selecting each field from the input records based on the result of the $\mathsf{pick}$ function. \Ruleref{syn} implements synthesis. As with analysis, the index $i$ of each entry in the generated record determines the needed inclusion map. %

It might seem like, once again, type information is used only in computing other type information. However, this is not quite true. The size of the right-hand side in each of rules \pref{rule:prj}, \pref{rule:concat}, and \pref{rule:syn} is determined by one of the input types. For \ruleref{concat}, the needed size is also the sum of the sizes of the input records. For \ruleref{prj}, it is the size (i.e., the size of the domain of definition) of the evidence. For \ruleref{syn}, however, there is seemingly no clue to the size of the required record other than the type.

Our experimental implementation erases all type information before evaluation. This requires delaying record synthesis until the size of the synthesized record is determined by another term, either a projection or concatenation. We leave formalizing the erased semantics, and proving it equivalent to the fully explicit semantics given here, for future work.

\subsection{Evidence}

\Cref{fig:red-evid} gives reduction rules for evidence. Reflexivity is interpreted as the identity inclusion map (\ruleref{leq-refl}). The combination of any row with the empty row is witnessed by the identity inclusion and the empty map (rules \pref{rule:plus-emptyl} and \pref{rule:plus-emptyr}). Transitivity is witnessed by composition of the individual maps (\ruleref{trans}). Transforming the types in rows does not change relationships between the rows (rules \pref{rule:leq-map} and \pref{rule:plus-map}). Rules (\ref{rule:leq-plusl}) and (\ref{rule:leq-plusr}) transform evidence for combination into evidence for containment; as we have chosen to represent combination with the corresponding containment maps, this is direct. Rules (\ref{rule:plus-compll}) and  (\ref{rule:plus-complr}) perform the dual transformation. In each case, we are provided only one of the two needed containment maps, and must compute the other. Finally, \ruleref{cong-evid} normalizes evidence in constant applications.

\begin{figure}
\begin{small}
\begin{minipage}[t]{0.44\linewidth}
\begin{align}
\LeqV {refl} &\RedQ
  \InclV \, (i \mapsto i)
\label{rule:leq-refl}
\tag{\rdrule{\LeqV {refl}}}
\\
\TransV {\InclV \, p} {\InclV \, q} &\RedQ
  \InclV \, (p \circ q)
\label{rule:trans}
\tag{\rdrule{\fatsemi}}
\\
\LeqV {map} \, (\InclV \, p) &\RedQ
  \InclV \, p
\label{rule:leq-map}
\tag{\rdrule{\LeqV {map}}}
\\
\LeqV {plusL} \, (\CombV \, p \, q) &\RedQ
  \InclV \, p
\label{rule:leq-plusl}
\tag{\rdrule{\LeqV {\mathord\odot L}}}
\\
\LeqV {plusR} \, (\CombV \, p \, q) &\RedQ
  \InclV \, q
\label{rule:leq-plusr}
\tag{\rdrule{\LeqV {\mathord\odot R}}}
\end{align}
\end{minipage}
\hspace{0.02\linewidth}
\begin{minipage}[t]{0.52\linewidth}
\begin{align}
\PlusV {emptyL} &\RedQ
  \CombV \, \emptyset \, (i \mapsto i)
\label{rule:plus-emptyl}
\tag{\rdrule{\PlusV {\mathord\odot L}}}
\\
\PlusV {emptyR} &\RedQ
  \CombV \, (i \mapsto i) \, \emptyset
\label{rule:plus-emptyr}
\tag{\rdrule{\PlusV {\mathord\odot R}}}
\\
\PlusV {map} \, (\CombV \, p \, q) &\RedQ
  \CombV \, p \, q
\label{rule:plus-map}
\tag{\rdrule{\PlusV {map}}}
\\
\PlusV {complL} \, (\InclV \, q) &\RedQ
  \CombV \, (\mathsf{dual}(q)) \, q
\label{rule:plus-compll}
\tag{\rdrule{\PlusV {complL}}}
\\
\PlusV {complR} \, (\InclV \, p) &\RedQ
  \CombV \, p \, (\mathsf{dual}(p))
\label{rule:plus-complr}
\tag{\rdrule{\PlusV {complR}}}
\end{align}
\end{minipage}
\[
E[S \AppE {E_{\mathcal Q} [Q]}] \Red
  E[S \AppE {E_{\mathcal Q} [Q']}]
  \qquad \qquad \text{if $Q \RedQ Q'$}
\label{rule:cong-evid}
\tag{$\xi_{\mathcal Q}$}
\]
\end{small}
\caption{Reduction: Evidence}
\label{fig:red-evid}
\end{figure}

\subsection{Properties}

\begin{lemma}[Substitution] ~
  \begin{itemize}
  \item If $\KindJ {\Delta, \alpha \co \kappa'} \tau \kappa$ and $\KindJ \Delta \upsilon {\kappa'}$, then $\KindJ \Delta {\tau[\upsilon/\alpha]} {\kappa}$\footnote{\CiteAgda{Rome.Types.Substitution}{\_$\beta_{k}[\_]$}}
  \item If $\EntJ {\Delta, \alpha \co \kappa} \Psi Q \pi$ and $\KindJ \Delta \tau \kappa$, then $\EntJ \Delta {\Psi[\tau/\alpha]} {Q[\tau/\alpha]} {\pi[\tau/\alpha]}$\footnote{\CiteAgda{Rome.Types.Substitution}{subPred$_{k}$}}
  \item If $\EntJ \Delta {\Psi, v \co \pi'} Q \pi$ and $\EntJ \Delta \Psi {Q'} {\pi'}$, then $\EntJ \Delta \Psi {Q[Q'/v]} \pi$.\footnote{\CiteAgda{Terms.Normal.Substitution}{subEnt}}
  \item If $\TypeJ {\Delta, \alpha\co\kappa} \Psi \Gamma M \tau$ and $\TypeJ \Delta \upsilon \kappa$ then $\TypeJ \Delta {\Psi[\upsilon/\alpha]} {\Gamma[\upsilon/\alpha]} {M[\upsilon/\alpha]} {\tau[\upsilon/\alpha]}$.\footnote{\CiteAgda{Terms.Normal.Substitution}{\_$\beta\cdot$[\_]}}
  \item If $\TypeJ \Delta {\Psi, v \co \pi} \Gamma M \tau$ and $\EntJ \Delta \Psi Q \pi$, then $\TypeJ \Delta \Psi \Gamma {M[Q/v]} \tau$.\footnote{\CiteAgda{Terms.Normal.Substitution}{\_$\beta\pi$[\_]}}
  \item If $\TypeJ \Delta \Psi {\Gamma, x \co \upsilon} M \tau$ and $\TypeJ \Delta \Psi \Gamma N \upsilon$, then $\TypeJ \Delta \Psi \Gamma {M[N/x]} \tau$.\footnote{\CiteAgda{Terms.Normal.Substitution}{\_$\beta$[\_]}}
  \end{itemize}
\end{lemma}

\begin{theorem}[Preservation] ~
  \begin{itemize}
  \item If $\KindJ {\Delta} \tau \kappa$ and $\tau \RedT \tau'$, then $\KindJ {\Delta} {\tau'} \kappa$.\footnote{The equivalence relation \CiteAgda{Rome.Types.Equivalence.Relation}{\_$\equiv$t\_} is intrinsically type-preserving}
  \item If $\EntJ {\Delta} {\Phi} Q \pi$ and $Q \RedQ Q'$ then $\EntJ {\Delta} {\Phi} {Q'} \pi$.\footnote{The reduction relation \CiteAgda{Terms.Normal.Reduction}{\_=$\Rightarrow$\_} is intrinsically type-preserving}
  \item If $\TypeJ {\Delta} {\Phi} {\Gamma} M \tau$ and $M \Red M'$ then $\TypeJ {\Delta} {\Phi} {\Gamma} {M'} \tau$.\footnote{The reduction relation \CiteAgda{Terms.Normal.Reduction}{\_--$\to$\_} is intrinsically type-preserving}
  \end{itemize}
\end{theorem}

\begin{theorem}[Progress] ~
  \begin{itemize}
  \item If $\EntJ {\varepsilon} {\varepsilon} Q \pi$ then either there is some $Q'$ such that $Q \RedQ Q'$, or $Q \in \NormalEvidence$.\footnote{\CiteAgda{Terms.Theorems.Progress}{entProgress}}
  \item If $\TypeJ {\varepsilon} {\varepsilon} {\varepsilon} M \tau$ then either there is some $M'$ such that $M \Red M'$, or $M \in \NormalTerms$.\footnote{\CiteAgda{Terms.Theorems.Progress}{progress}}
  \end{itemize}
\end{theorem}

\section{Related Work}
\label{sec:related}

\newcommand\Moo[1]{\mathrm{Mu}\, #1}

We conclude with a discussion of related and future work. There is a significant literature on row types and on extensible data types in general.  We highlight the work that is most relevant to \Rome.

\subsubsection*{Featherweight Ur.}

The most similar language to \Rome is Featherweight Ur~\citep{Chlipala10}.  Ur also supports row and record concatenation with first-class labels, enabled by first-class label inequality proofs. Ur is based on System~\FO, and supports mapping type-level operations over rows. In contrast to \Rome, Ur is focused on recursive records, and their data-driven unfolding. Ur does not share \Rome's focus on recursive variants, or on the duality between records and variants.

\subsubsection*{Polymorphic variants in OCaml}

\InlineOn
Extensible recursive variants exist in OCaml~\citep{OCamlObjTypes,OCamlVariants}, as introduced (and later formalized \citep{Garrigue10}) by \citet{Garrigue98}. Rather than using row types, OCaml captures relationships between extensible variant types using subtyping. Similar to the extended expression problem, \citet{Garrigue00} uses polymorphic variants to encode modular evaluation. They do not explicitly consider mixed return types, such as our !evalA | evalB! \cref{sec:big-step}, but we have succeeded in encoding these examples in OCaml\Supplemental. Due to the lack of first-class labels, label kinds, and row kinds, the types of these OCaml analogues lack some precision. Additionally, our first-class label combinators (\cref{fig:fcl}) are impossible to express. This has application to recursive programming: consider a higher-order version of !dnaw! \cref{sec:variants-and-records}, which can be seen as selecting a particular case handler from two $F$-algebras. Quantification of label variables make this type impossible to express in OCaml, while the use of branching makes its term impossible to define.

\begin{rosi}
dnawHO : forall x y z t u l. x + y ~ z, {l := t} < z =>
         (Sigma x u -> u) -> (Sigma y u -> u) -> #l -> t u -> u
dnawHO = \ f g l x. (f | g) (inj (l := x))
\end{rosi}

\noindent
The differences between our approaches are also illustrated by the desugaring example \cref{sec:desugaring}. OCaml does not have an analogue of our label-generic abstractions, and so combining generic traversals still requires some enumeration of cases. More significantly, however, the result of desugaring in OCaml does not have the precision that it does in \Rome: our type guarantees that the Boolean cases are removed, whereas the result in OCaml permits Boolean cases to still be present.

\subsubsection*{Encodings rows} Several encodings of row types have been explored in other type systems, notably Haskell's type classes and type families \citep{Bahr14, KiselyovLS04, Morris15, OliveiraMY15, Swierstra08}. These encodings often represent rows as specific sequences of types and struggle to capture the full flexibility intended by row typing. As an example, coproducts of functors in \citeposs{Swierstra08} encoding must be explicitly right-nested to for typeclass instances to resolve---in essence, the encoding fails to capture the commutativity and associativity of rows. Hence we advocate for row types as first-class citizens of the type theory.

\subsubsection*{Intersection types and the merge operator}
Extensible data types can be directly expressed using intersection types and the merge operator \citep{Dunfield12}. Albeit expressive, the merge operator can be a difficult beast to tame safely~\citep{HuangZO21, RiouxHOZ23}. Our work differs in approach: we advocate for the more restricted containment relation on rows $\LeqP {\rho_{1}} {\rho_{2}}$ as opposed to full-fledged subtyping. A comparison can be drawn between our complement operator and the \emph{type difference} operator of \citeposs{XuHO23} intersection type system, which generalizes record complements to arbitrary type. Again, the two differ in approach, as the type difference operator is strictly more general.

\subsubsection*{Intrinsic mechanization \& normalization by evaluation in Agda}
Our metatheoretic results and mechanization owe a great debt to \citet{ChapmanKNW19}, who in turn cite \citet{AllaisMB13, AltenKirchR99} in the design of their intrinsic syntax, normalization algorithm, and metatheory. Likewise, we follow \citeposs{AllaisMB13} formalization of lists in finding the correct semantic image and normal form of rows. The overall structure of proof that they (and we) employ is otherwise standard for normalization by evaluation techniques (\cf{}~\citet{Lindley05}). %

\section{Future Work}
\label{sec:future}

\subsubsection*{Erasure}
\label{sec:erasure}

Our operational semantics relies on type information in the interpretation of the \lstinline!syn! primitive. Of course, a more satisfying account would avoid any runtime reliance on types. This would be both practically beneficial, as it would guarantee that type information could be erased, and semantically preferable, as it would suggest that \Rome could enjoy a parametricity theorem. Our experimental implementation does erase type information before execution, relying on a more subtle implementation of \lstinline!syn! (and of evidence for the complement rules). We hope to formalize this erased semantics, and show that it simulates the fully type-annotated semantics presented here.

\subsubsection*{Relating semantics}

We give an operational semantics of \Rome in Agda, whereas \citet{HubersM23} provide a denotational shallow embedding. \citet{SaffrichTM24} demonstrates that big-step operational semantics and denotational semantics of System $F$ can be related in Agda using a step-indexed logical relation. We hope to adopt their technique to characterize a relation between the denotational and operational accounts of \RO and \Rome.

\ifreview
\else
\section*{Data Availability Statement}
\label{sec:data}
Our Agda mechanization of \Rome is available online \citep{artifactRome}.

\begin{acks}

\end{acks}
\fi

\bibliographystyle{ACM-Reference-Format}
\bibliography{rome}

\ifextended

\clearpage

\InlineOff
\subfile{Appendix.tex}

\fi

\end{document}

%% file: Appendix.tex
\appendix

\section{Omitted Typing Rules}
\label{sec:extra-typing-rules}

\subsubsection*{Environment formation.}

\begin{small}
\begin{gather*}
\fbox{$\PEnvJ \Delta \Phi$} \; \; \fbox{$\TEnvJ \Delta \Gamma$}
\\
\begin{gathered}
\ib{\irule[\crule{pemp}]
          { };
          {\PEnvJ \Delta \varepsilon}}
\rsp
\ib{\irule[\crule{pvar}]
          {\PEnvJ \Delta \Phi}
          {\PredJ \Delta \pi};
          {\PEnvJ \Delta {\Phi, v : \pi}}}
\end{gathered}
\rsp
\begin{gathered}
\ib{\irule[\crule{temp}]
          { };
          {\TEnvJ \Delta \varepsilon}}
\rsp
\ib{\irule[\crule{tvar}]
          {\TEnvJ \Delta \Gamma}
          {\KindJ \Delta \tau \TypeK};
          {\TEnvJ \Delta {\Gamma, x : \tau}}}
\end{gathered}
\end{gather*}
\end{small}

\subsubsection*{Type equivalence.}

\begin{small}
\begin{gather*}
\fbox{$\TEqvJ \Delta \tau \tau \kappa$}
\\
\ib{\irule[\erule{refl}]
          {\KindJ \Delta \tau \kappa};
          {\TEqvJ \Delta \tau \tau \kappa}}
\rsp
\ib{\irule[\erule{sym}]
          {\TEqvJ \Delta {\tau_1} {\tau_2} \kappa};
          {\TEqvJ \Delta {\tau_2} {\tau_1} \kappa}}
\rsp
\ib{\irule[\erule{trans}]
          {\TEqvJ \Delta {\tau_1} {\tau_2} \kappa}
          {\TEqvJ \Delta {\tau_2} {\tau_3} \kappa};
          {\TEqvJ \Delta {\tau_1} {\tau_3} \kappa}}
\\
\ib{\irule[\erule{$\eta$}]
          {\KindJ \Delta {\phi} {\kappa_1 \to \kappa_2}};
          {\TEqvJ \Delta {\phi} {\lambda \alpha\co\kappa_1. \phi \, \alpha} {\kappa_1 \to \kappa_2}}}
\rsp
\ib{\irule[\erulec{\Sing {}}]
          {\TEqvJ \Delta {\xi_1} {\xi_2} \kappa};
          {\TEqvJ \Delta {\Sing{\xi_1}} {\Sing{\xi_2}} \TypeK}}
\rsp
\ib{\irule[\erulec {\then}]
          {\PEqvJ \Delta {\pi_1} {\pi_2}}
          {\TEqvJ \Delta {\tau_1} {\tau_2} \TypeK};
          {\TEqvJ \Delta {\pi_1 \then \tau_1} {\pi_2 \then \tau_2} \TypeK}}
\\
\ib{\irule[\erulec{\text{\textsc{app}}}]
          {\TEqvJ \Delta {\tau_1} {\upsilon_1} {\kappa \to \kappa'}}
          {\TEqvJ \Delta {\tau_2} {\upsilon_2} {\kappa}};
          {\TEqvJ \Delta {\tau_1\,\tau_2} {\upsilon_1\,\upsilon_2} {\kappa'}}}
\rsp
\ib{\irule[\esrule{$\xi_{\{\,\}}$}]
          {\TEqvJ \Delta {\xi\One_i} {\xi\Two_i} \LabK}
          {\TEqvJ \Delta {\tau\One_i} {\tau\Two_i} \kappa};
          {\TEqvJ \Delta {\RowIx i 1 n {\LabTy {\xi\One_i} {\tau\One_i}}} {\RowIx i 1 n {\LabTy {\xi\Two_i} {\tau\Two_i}}} {\RowK\kappa}}}
\\
\ib{\irule[\erulec {\forall}]
          {\TEqvJ {\Delta, \gamma\co\kappa} {\tau[\gamma/\alpha]} {\upsilon[\gamma/\beta]} \TypeK};
          {\TEqvJ \Delta {\forall\alpha\co\kappa.\tau} {\forall\beta\co\kappa.\upsilon} \TypeK}}
\rsp
\ib{\irule[\erulec {\lambda}]
          {\TEqvJ {\Delta, \gamma\co\kappa} {\tau[\gamma/\alpha]} {\upsilon[\gamma/\beta]} {\kappa'}};
          {\TEqvJ \Delta {\lambda\alpha\co\kappa.\tau} {\lambda\beta\co\kappa.\upsilon} {\kappa \to \kappa'}}}
\rsp
\text{($\gamma$ fresh for $\tau, \upsilon$)}
\\
\ib{\irule[\emrulec{\mu}]
          {\TEqvJ \Delta {\phi_1} {\phi_2} {\kappa \to \kappa}};
          {\TEqvJ \Delta {\RecTy {\phi_1}} {\RecTy {\phi_2}} \kappa}}
\rsp
\ib{\irule[\errulec{\Xi}]
          {\TEqvJ \Delta {\rho_1} {\rho_2} {\RowK\kappa}};
          {\TEqvJ \Delta {\Xi \rho_1} {\Xi \rho_2} {\kappa}}!
          {\Xi \in \Set {\Pi, \Sigma}}}
\\
\ib{\irule[\erulec{\lesssim}]
          {\TEqvJ \Delta {\tau_i} {\upsilon_i} {\RowK\kappa}};
          {\PEqvJ \Delta {\LeqP {\tau_1} {\tau_2}} {\LeqP {\upsilon_1} {\upsilon_2}}}}
\rsp
\ib{\irule[\erulec{\odot}]
          {\TEqvJ \Delta {\tau_i} {\upsilon_i} {\RowK\kappa}};
          {\PEqvJ \Delta {\PlusP {\tau_1} {\tau_2} {\tau_3}} {\PlusP {\upsilon_1} {\upsilon_2} {\upsilon_3}}}}
\end{gather*}
\end{small}

\section{Omitted data type definitions}
\label{sec:extra-data-types}

\subsubsection*{Unit}~
\begin{rosi}
type Unit : * 
type Unit = #'Unit 

tt : Unit
tt = #'Unit
\end{rosi}
\subsubsection*{Tuples}~
\begin{rosi}
type Pair : * -> * -> *
type Pair = \ t u. Pi {'1 := t, '2 := u}

pair : forall t u. t -> u -> Pair t u
pair = \x y. (#'1 := x) ++ (#'2 := y)

type Triple : * -> * -> * -> *
type Triple = \ t u v. Pi {'1 := t, '2 := u, '3 := v}

triple : forall t u v. t -> u -> v -> Triple t u v
triple = \x y z. pair x y ++ (#'3 := z)

fst = \x. sel x #'1
snd = \x. sel x #'2
thd = \x. sel x #'3
\end{rosi}

\subsubsection*{Naturals}~

\begin{rosi}
type NatF : * -> *
type NatF = \n. Sigma { 'Zero := Unit, 'Succ := n}

type Nat : *
type Nat = Mu NatF

zero : Nat
zero = in (inj (#'Zero := tt))

succ : Nat -> Nat
succ = \n. in (inj (#'Succ := n))

add : Nat -> Nat -> Nat
add = \m. fix (\add n. 
                 (case #'Zero (\u. m) 
               | case #'Succ (\nn. succ (add nn))) (out n))

one : Nat
one = succ zero

two : Nat
two = succ one

three : Nat
three = add one two
\end{rosi}

\subsubsection*{Lists}~
\begin{rosi}
type ListF : * -> * -> *
type ListF = \a. Sigma { 'Nil := Const Unit, 'Cons := (\l. Pair a l) }

type List : * -> *
type List = \a. Mu (ListF a)

nil : forall a. List a
nil = rcon #'Nil tt

cons : forall a. a -> List a -> List a
cons = \hd tl. rcon #'Cons (pair hd tl)

head : forall a. List a -> Maybe a
head = \l. match (out l)
           ( case #'Nil (const Nothing)
           | case #'Cons (o Just fst)
           )

tail : forall a. List a -> List a
tail = \l. match (out l)
           ( case #'Nil (const nil)
           | case #'Cons snd
           )

nth : forall a. List a -> Nat -> Maybe a
nth = fix (\f l n.
            match (out n)
            ( case #'Zero (const (head l))
            | case #'Succ (f (tail l))
            ))
\end{rosi}

\section{Omitted Code Samples}
\label{sec:extra-code-samples}

\subsubsection*{Evaluation: Boolean language}~

\begin{rosi}
type ValB : R[R[* -> *] -> * -> *]
type ValB = { 'Bool := (\ expr. \ val. Bool), 'Err := (\ expr. \ val. Unit) }
\end{rosi}
\codesep
\begin{rosi}
evalB : forall valr. ValB < valr =>
        Xh BoolF (\ expr. Env (valr expr) -> Mu (Sigma (valr expr)))
evalB = \ exp rec env.
           match exp
           ( case #'BConst (rcon #'Bool)
           | case #'If (\ t.
               as #'Bool (rec (sel t #'1) env)
               ( case #'True (const (rec (sel t #'2) env))
               | case #'False (const (rec (sel t #'3) env)))))
\end{rosi}

\section{Omitted Evaluation Definitions}
\label{sec:extra-evaluation}

\subsubsection*{The $pick$ function}

\begin{align*}
  \mathsf{pick} &: (\mathbb N \pto \mathbb N) \times (\mathbb N \pto \mathbb N) \times \mathbb N \pto \mathbb N + \mathbb N \\
  \mathsf{pick}(p, q, i) &= \begin{cases}
    \mathsf{Left} \, j &\text{if $p(j) = i$} \\
    \mathsf{Right} \, j &\text{if $q(j) = i$}
  \end{cases}
\end{align*}

\subsubsection*{The $dual$ function}

\[
\begin{aligned}
  \mathsf{dual} &: (\mathbb N \pto \mathbb N) \to \mathbb N \pto \mathbb N \\
  \mathsf{dual}(p, i) &= \mathsf{go}_p(i, 0)
\end{aligned}
\qquad\qquad
\begin{aligned}
  \mathsf{go}_p &: \mathbb N \pto \mathbb N \pto \mathbb N \\
  \mathsf{go}_p(0, j) &=
    \begin{cases}
      j &\text{if $j \not\in \ran(p)$} \\
      \mathsf{go}_p(0, j + 1) &\text{otherwise}
    \end{cases} \\
  \mathsf{go}_p(i, j) &=
    \begin{cases}
      \mathsf{go}_p(i, j + 1) &\text{if $j \in \ran(p)$} \\
      \mathsf{go}_p(i - 1, j + 1) &\text{otherwise}
    \end{cases}
\end{aligned}
\]